\documentclass[longauth]{aa}
\usepackage{graphicx}
\usepackage{natbib}
\titlerunning{}

\def\te{T_{\rm eff}}
\def\Mo{M_{\odot}}
\def\Lo{L_{\odot}}

\usepackage{color}

\newcommand{\kms}{\mbox{$\mbox{km\,s}^{-1}$}\,}
\newcommand{\ms}{\mbox{$\mbox{m\,s}^{-1}$}\,}

\begin{document}

\title{HARPS-N high spectral resolution observations of Cepheids \\ I. The Baade-Wesselink projection factor of $\delta$~Cep revisited\thanks{Table A.1 is available in electronic format at the CDS via anonymous ftp to cdsarc.u-strasbg.fr (130.79.128.5) or via http://cdsarc.u-strasbg.fr/viz-bin/qcat?J/A+A/593/A45}}
\titlerunning{Revisiting The Baade-Wesselink projection factor of $\delta$~Cep}
\authorrunning{Nardetto et al. }

\author{N.~Nardetto \inst{1} 
\and E.~Poretti \inst{2} 
\and M.~Rainer \inst{2} 
\and A.~Fokin \inst{3}
\and P.~Mathias\inst{4, 5}
\and R.I.~Anderson \inst{6}
\and A.~Gallenne\inst{7,8}  
\and W.~Gieren\inst{8, 9}  
\and D.~Graczyk \inst{8, 9, 10} 
\and P.~Kervella \inst{11,12}  
\and A.~M\'erand \inst{7} 
\and D.~Mourard \inst{1}  
\and H.~Neilson\inst{13} 
\and G.~Pietrzynski \inst{10}  
\and B.~Pilecki \inst{10}  
\and J.~Storm \inst{14}  
}
\institute{Universit\'e C\^ote d'Azur, OCA, CNRS, Lagrange, France,  Nicolas.Nardetto@oca.eu   
\and  INAF -- Osservatorio Astronomico di Brera, Via E. Bianchi 46, 23807 Merate (LC), Italy 
\and Institute of Astronomy of the Russian Academy of Sciences, 48 Pjatnitskaya Str., 109017, Moscow, Russia 
 \and   Universit\'e de Toulouse, UPS-OMP, Institut de Recherche en Astrophysique et Plan\'etologie, Toulouse, France
\and CNRS, UMR5277,  Institut de recherche en Astrophysique et Plan\'etologie, 14 Avenue Edouard Belin, 31400 Toulouse, France 
\and Department of Physics and Astronomy, The Johns Hopkins University, 3400 N. Charles St, Baltimore, MD, 21218, USA
\and European Southern Observatory, Alonso de C\'ordova 3107, Casilla 19001, Santiago 19, Chile 
\and Departamento de Astronom\'ia, Universidad de Concepci\'on, Casilla 160-C, Concepci\'on, Chile 
\and Millenium Institute of Astrophysics, Santiago, Chile 
\and Nicolaus Copernicus Astronomical Center, Polish Academy of Sciences, ul. Bartycka 18, PL-00-716 Warszawa, Poland
\and LESIA (UMR 8109), Observatoire de Paris, PSL, CNRS, UPMC, Univ. Paris-Diderot, 5 place Jules Janssen, 92195 Meudon, France 
\and Unidad Mixta Internacional Franco-Chilena de Astronom\'ia, CNRS/INSU, France (UMI 3386) and Departamento de Astronom\'ia, Universidad de Chile, Camino El Observatorio 1515, Las Condes, Santiago, Chile 
\and Department of Astronomy \& Astrophysics, University of Toronto, 50 St. George Street, Toronto, ON, M5S 3H4 
\and Leibniz Institute for Astrophysics, An der Sternwarte 16, 14482, Potsdam, Germany 
}

\date{Received ... ; accepted ...}

\abstract{The projection factor $p$ is the key quantity used in the Baade-Wesselink (BW) method for  distance determination; it converts radial velocities into pulsation velocities.  Several methods are used to determine $p$, such as geometrical and hydrodynamical models or the inverse BW approach when the distance is known. }{We analyze new HARPS-N spectra of $\delta$~Cep to measure its cycle-averaged atmospheric velocity gradient in order to better constrain the projection factor.} {We first apply the inverse BW method to derive $p$ directly from observations. The projection factor can be  divided into three subconcepts: (1) a geometrical effect ($p_\mathrm{0}$), (2) the velocity gradient within the atmosphere ($f_\mathrm{grad}$), and (3) the relative motion of the optical pulsating photosphere with respect to the corresponding mass elements ($f_\mathrm{o-g}$).  We then measure the $f_\mathrm{grad}$ value of $\delta$ Cep for the first time.} {When the HARPS-N mean cross-correlated line-profiles are fitted with a Gaussian profile, the projection factor is $p_\mathrm{cc-g}=1.239\pm0.034 (stat.)\pm0.023 (syst.)$. When we consider the different amplitudes of the radial velocity curves that are associated with 17 selected spectral lines, we measure projection factors ranging from 1.273 to 1.329. We find a relation between $f_{\rm grad}$ and the line depth measured when the Cepheid is at minimum radius. This relation is consistent with that obtained from our best hydrodynamical model of $\delta$~Cep and with our projection factor decomposition. Using the observational values of $p$ and $f_\mathrm{grad}$ found for the 17 spectral lines, we derive a semi-theoretical value of $f_\mathrm{o-g}$. We alternatively obtain $f_\mathrm{o-g}=0.975 \pm 0.002$ or $1.006 \pm 0.002$ assuming models using radiative transfer in plane-parallel or spherically symmetric geometries, respectively.} {The new HARPS-N observations of $\delta$~Cep are consistent with our decomposition of the projection factor. The next step will be to measure  $p_\mathrm{0}$ directly from the next generation of visible interferometers. With these values in hand, it will be possible to derive  $f_\mathrm{o-g}$ directly from observations.}
\keywords{Techniques: interferometry, spectroscopy -- Stars: circumstellar matter -- Stars: oscillations (including pulsations) -- Stars individual: $\delta$~Cep}
\maketitle

\section{Introduction}\label{s_Introduction}

Since their period-luminosity (PL) relation was established \citep{leavitt1912}, Cepheid variable stars have been used to calibrate the distance scale \citep{hertzsprung13} and then the Hubble constant \citep{riess11, freedman12, riess16}. The discovery that the $K$-band PL relation is nearly universal and can be applied to any host galaxy whatever its metallicity \citep{storm11a} is a considerable step forward in the use of Cepheids as distance indicators.
 Determining the distances to Cepheids relies on the Baade-Wesselink (BW) method, which in turn relies on a correct evaluation of the projection factor $p$. This is necessary to convert the radial velocity variations derived from the spectral line profiles into photospheric pulsation velocities \citep{nardetto04}.                                                                                             

The projection factor of $\delta$~Cep, the eponym of the Cepheid variables, has been determined by means of different techniques, which we summarize here (see Table~\ref{tab_history} and the previous review by \citealt{nardetto14b}):

\begin{itemize}
\item
Purely geometric considerations lead to the identification of two contributing effects only in the projection factor, i.e., the 
limb darkening of the star and the motion (expansion or contraction) of the atmosphere. \citet{nardetto14b} described this classical approach and its recent variations \citep[e.g., ][]{gray07,hadrava09b}.


\begin{table*}
\begin{center}
\caption{\label{tab_history} Non-exhaustive history of the determination of the Baade-Wesselink projection factor in the case of $\delta$~Cep. The method used to derive the radial velocity is indicated, and cc-g corresponds to a Gaussian fit of the cross-correlated line profile. For the values of the projection factor derived from a published period projection factor relation, we  consistently use a period of $P=5.366208$ days \citep{engle14}.}
\setlength{\tabcolsep}{4pt}
\begin{tabular}{|l|l||l|}
\hline
method &  $p$ & reference \\
\hline
\multicolumn{3}{|c|}{Geometrical models}  \\
\hline
  centroid & 1.415 & \citet{getting34} \\
    centroid & 1.375   &\citet{vanhoof52}\\
      centroid & 1.360  &\citet{burki82} \\
      centroid & 1.328  &\citet{neilson12} \\
      \hline
\multicolumn{3}{|c|}{Hydrodynamical models}  \\
\hline
  bisector & 1.34 & \cite{sabbey95} \\
    Gaussian &1.27 $\pm$ 0.01  & \citet{nardetto04} \\
       cc-g ($Pp$) & 1.25 $ \pm$ 0.05   &\citet{nardetto09}  \\
       \hline
\multicolumn{3}{|c|}{Observations}  \\
\hline
     cc-g & 1.273 $\pm$ 0.021 $\pm$ 0.050 &\citet{merand05} \\
     cc-g & 1.245 $\pm$ 0.030 $\pm$ 0.050 &\citet{gro07}  \\
     cc-g &1.290 $\pm$ 0.020 $\pm$ 0.050 &\citet{merand15} \\
\hline    
     cc-g ($Pp$)  & 1.47 $\pm$ 0.05  &\citet{gieren05}  \\
     cc-g  ($Pp$)  &1.29 $\pm$ 0.06  &\citet{laney09} \\
     cc-g  ($Pp$) &1.41 $\pm$ 0.05  &\citet{storm11b} \\
     cc-g   ($Pp$)  &1.325 $\pm$ 0.03   &\citet{gro13}  \\   
\hline
\end{tabular}
\end{center}
\end{table*}


\item To improve the previous method we should consider that  Cepheids do not pulsate in a quasi-hydrostatic way and the dynamical structure of their atmosphere is extremely complex \citep{sanford56,bell64, karp75, sasselov90,  wallerstein15}. Therefore,  improving the determination of the BW projection factor
requires a hydrodynamical model that is able to describe the atmosphere. To date, the projection factor has been studied with two such models: the first is based on a {\it piston} in which the radial velocity curve is used as an input \citep{sabbey95} and the second is a {\it self-consistent}  model \citep{nardetto04}. \citet{sabbey95} found a mean value of the projection factor  $p=1.34$ (see also \citealp{marengo02, marengo03}). However, this value was derived using the bisector method of the radial velocity determination (applied to the theoretical line profiles). This makes it difficult to compare this value with other studies. As commonly done in the literature, if a Gaussian fit of the cross-correlated line-profile is used to derive the radial velocity RV$_\mathrm{cc-g}$ (``cc'' for cross-correlated and ``g'' for Gaussian), then the measured projection factor tends to be about 11\% smaller than  the initial geometrical projection factor is found, i.e., $p=1.25\pm0.05$ \citep{nardetto09}. 
\item As an approach entirely based on observations, \citet{merand05} applied the inverse BW method to  infrared interferometric data of $\delta$~Cep. The projection factor is then fit, where the distance of $\delta$~Cep is known with 4\% uncertainty from the HST parallax ($d=274\pm11$~pc;  \citealt{benedict02}). They found $p=1.273\pm0.021\pm0.050$ using RV$_\mathrm{cc-g}$ for the radial velocity.  The first error is the internal one  due to the fitting method. The second is due to the uncertainty of the distance. \citet{gro07} found $p=1.245\pm0.030\pm0.050$ when using almost the same distance (273 instead of 274~pc), a different radial velocity dataset and a different fitting method of the radial velocity curve. Recently, \citet{merand15} applied an integrated inverse method (SPIPS) to $\delta$~Cep (by combining interferometry and photometry) and found $p=1.29\pm0.02\pm0.05.$ These values  agree closely with the self-consistent hydrodynamical model. Another slightly different approach is to apply the infrared surface brightness inverse method to distant Cepheids (see \citealp{fouque97, kervella04b} for the principles) in order to derive a period projection factor relation ($Pp$). In this approach the distance to each LMC Cepheid  is assumed to be the same by taking into account the geometry of LMC. This constrains the slope of the $Pp$ relation. However, its zero-point is alternatively  fixed using  distances to  Cepheids in Galactic clusters \citep{gieren05}, HST parallaxes of nearby Cepheids derived by \citet{vl07} \citep{laney09, storm11b}, or a combination of both \citep{gro13}.  \citet{laney09} also used  high-amplitude $\delta$ Scuti stars to derive their $Pp$ relation (see also Fig. 10 in  \citealp{nardetto14}). The projection factors derived by \citet{laney09} and \citet{gro13} are consistent with the interferometric values, while the projection factors from  \citet{gieren05} and \citet{storm11b} are significantly greater (see Table~\ref{tab_history}).

\item \citet{pilecki13} 
constrained the projection factor using a short-period Cepheid ($P=3.80$~d, similar to the period of $\delta$~Cep) in a eclipsing binary system. They found $p=1.21\pm0.04$. 
\end{itemize}

This non-exhaustive review shows just how complex  the situation regarding the value of the BW projection factor of $\delta$~Cep is. 
This paper is part of the international ``Araucaria~Project'', whose purpose is to provide an improved local calibration of the extragalactic distance scale out to distances of a few megaparsecs~\citep{gieren05_messenger}. 
In Sect.~\ref{s_HARPS-N} we present new High Accuracy Radial velocity Planet Searcher for the Northern hemisphere (HARPS-N) observations. 
Using these spectra together with the \citet{merand05} data obtained with the Fiber Linked Unit for Optical Recombination \citep[FLUOR, ][] {foresto97} operating at the focus of the Center for High Angular Resolution Astronomy (CHARA) array \citep{ten05} located at the Mount Wilson Observatory (California, USA), we apply the inverse BW method to derive the projection factor associated with the RV$_\mathrm{cc-g}$ radial velocity and for 17 individual spectral lines (Sect.~\ref{s_BWinv}). In Sect.~\ref{s_hydro}, we briefly describe the hydrodynamical model used in \citet{nardetto04} and review the projection factor decomposition into three sub-concepts ($p =  p_\mathrm{0} f_\mathrm{grad} f_\mathrm{o-g}$, \citealt{nardetto07}). In Sect.~\ref{ss_BWinv_model}, we compare the observational and theoretical projection factors. We then compare the hydrodynamical model with  the observed atmospheric velocity gradient (Sect.~\ref{ss_grad}) and the angular diameter curve of FLUOR/CHARA (Sect.~\ref{ss_RV}). In Sect.~\ref{s_fog} we derive the $f_\mathrm{o-g}$ quantity from the previous sections. 
We conclude in Sect.~\ref{s_conclusion}. 

\section{HARPS-N spectroscopic observations}\label{s_HARPS-N}

\begin{figure}[htbp]
\begin{center}
\resizebox{1.0\hsize}{!}{\includegraphics[clip=true]{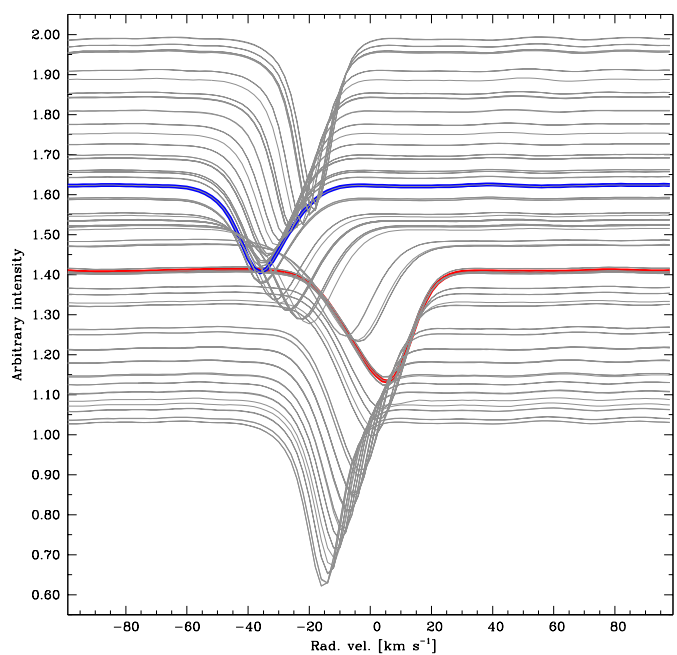}}
\end{center}
\caption{Mean line profile changes during the pulsation cycle of $\delta$~Cep.  The line profile with the highest receding
motion is highlighted in red, the one with the highest approaching motion in blue.}
\label{f0}
\end{figure}

HARPS-N is a high-precision radial-velocity spectrograph installed at the Italian Telescopio Nazionale Galileo (TNG), a 3.58-meter telescope located at the Roque de los Muchachos Observatory on the island of La Palma, Canary Islands, Spain \citep{co12}.  HARPS-N is the northern hemisphere counterpart of the similar HARPS instrument installed at the ESO 3.6 m telescope at La Silla Observatory in Chile. The instrument covers the wavelength range from 3800 to 6900~\AA\ with a resolving power of $R \simeq 115000$. A total of 103 spectra were secured between 27 March and 6 September 2015 in the framework of the OPTICON proposal 2015B/015 (Table~\ref{Tab_log}). In order to calculate the pulsation phase of each spectrum, we used
$P=5.366208$~d \citep{engle14} and $T_0=2457105.930$~d, the time
corresponding to the maximum approaching velocity determined from the HARPS-N radial velocities.
The data are spread over 14 of the 30 pulsation cycles  elapsed between the first and last epoch.
The final products of the HARPS-N data reduction software (DRS) installed at TNG (on-line mode)
are background-subtracted, cosmic-corrected, flat-fielded, and wavelength-calibrated spectra
(with and without merging of the spectral orders).
We used these spectra to compute the mean-line profiles by means of the least-squares deconvolution
(LSD) technique \citep{donati97}. As can be seen in
Fig.~\ref{f0}, the mean profiles reflect the large-amplitude radial pulsation by distorting
the shape in a continuous way.

\begin{figure*}[htbp]
\begin{center}
\resizebox{0.8\hsize}{!}{\includegraphics[clip=true]{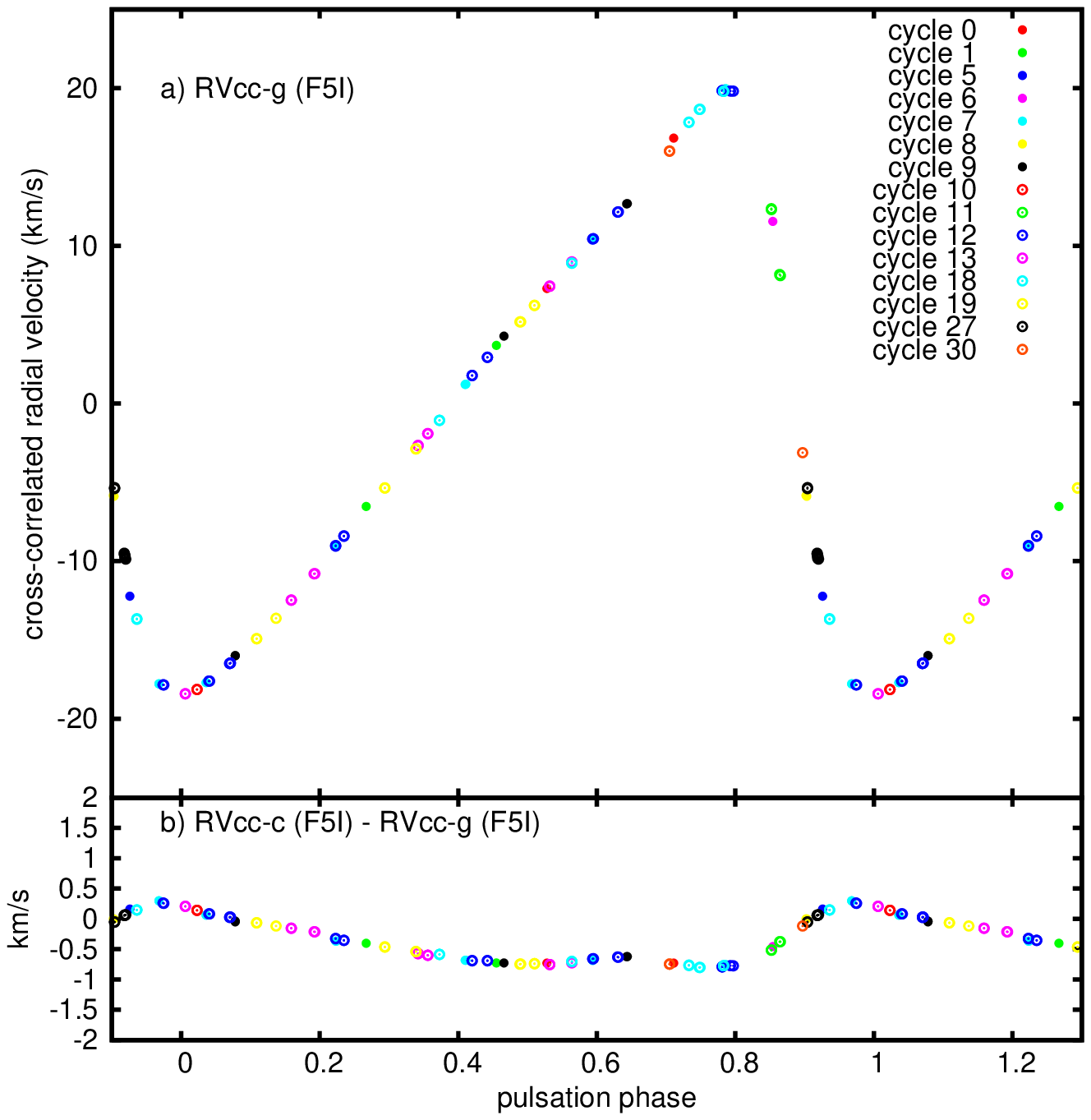}}
\end{center}
\caption{In panel~a), the HARPS-N radial velocities associated with the Gaussian fit of the cross-correlated line profile (RV$_\mathrm{cc-g}$ using F5I template) are plotted as a function of  the pulsation phase of the star after correction of the $\gamma$-velocity (i.e., $V_\mathrm{\gamma}= -16.95$ \kms). The cycle of observations are shown in different colors. The data are reproduced from one cycle to the other. The precision on the measurements is between 0.5 and 1.5 \ms (error bars are lower than symbols). In panel~b) we show  the residuals between RV$_\mathrm{cc-c}$ (F5I) and RV$_\mathrm{cc-g}$ (F5I) after correction of their respective $\gamma$-velocities.} \label{f1a_RVcc}
\end{figure*}

The DRS also computes the star's radial velocity by
fitting a Gaussian function to the cross-correlation functions (CCFs). To do this,
the DRS uses a mask including thousands of lines
covering the whole HARPS-N spectral ranges. The observer can  select the mask among
those available online and the G2 mask was the closest to the $\delta$~Cep spectral type.
As a further step, we re-computed the RV values  by using the HARPS-N
DRS in the offline mode on the Yabi platform, as in the case of $\tau$~Boo  \citep{borsa15}.
We applied both
library and custom masks, ranging spectral types from F5 to G2.
We obtained RV curves fitted with very similar sets of least-squares parameters 
showing only some small changes in the mean values. Table~\ref{Tab_log}
lists the $\mathrm{RV_\mathrm{cc-g}}$ values obtained from the custom F5I mask,
also plotted in  panel a) of Fig.~\ref{f1a_RVcc} with different symbols
for each pulsation cycle. The RV curve shows a
full amplitude of  38.4~\kms\, and an average value (i.e., the A$_0$ value
of the fit) of $V_\gamma =-16.95$~\kms.
As a comparison, we calculate the centroid of the mean line
profiles $\mathrm{RV_\mathrm{cc-c}}$ (``cc'' for cross-correlated
and ``c'' for centroid) and plot the residuals in panel b)
of Fig.~\ref{f1a_RVcc}. The $\mathrm{RV_\mathrm{cc-c}}$
curve has an amplitude 1.2~\kms\, lower than that of
the $\mathrm{RV_\mathrm{cc-g}}$ value. This result is similar to that
reported in the case of $\beta$~Dor \citep[Fig.~2 in ][]{nardetto06a}.
The implications of this difference on the projection factor is
discussed in Sect.~\ref{ss_RVcc}.

Consistent with \citet{anderson15a}, we find no
evidence for cycle-to-cycle differences in the RV amplitude
as exhibited by long-period Cepheids \citep{anderson14,anderson16}.
We also investigated the possible effect of the binary motion due to the
companion \citep{anderson15a}. Unfortunately, the HARPS-N
observations are placed on the slow decline of the RV curve and they
only  span 160 days. We were able to  detect a slow drift of
$-0.5\pm0.1$~m~s$^{-1}$~d$^{-1}$. The least-squares solution with 14 harmonics leaves
a r.m.s. residual  of 49~m~s$^{-1}$.
Including  a linear trend did not significantly
reduce the residual since the major source of error
probably lies in the fits of the very different shapes of the mean-line profiles along
the pulsation cycle (Fig.~\ref{f0}).
We also note  that  surface effects induced by convection and granulation
\citep{neilson14}  could  contribute to increasing the residual r.m.s. These effects are also
observed in the light curves  of Cepheids \citep{derekas12,evans15,poretti15,derekas16}.

\begin{table*}
\begin{center}
\caption{HARPS-N spectroscopic data of $\delta$~Cep.} \label{Tab_log}
\setlength{\doublerulesep}{\arrayrulewidth}
 \tiny
\begin{tabular}{|lcrl|lcrl|}
\hline \hline \noalign{\smallskip}

BJD     &       $\phi$  &       RV$_\mathrm{cc-g}$      &       $\sigma_\mathrm{RV_\mathrm{cc-g}}$      &       BJD     &       $\phi$  &       RV$_\mathrm{cc-g}$      &       $\sigma_\mathrm{RV_\mathrm{cc-g}}$      \\

\hline
2457108.765     &       0.53    &       -9.6505 &       0.0004  &       2457174.601     &       0.80    &       2.8618  &       0.0009  \\
2457109.760     &       0.71    &       -0.1170 &       0.0005  &       2457175.553     &       0.97    &       -34.7911        &       0.0010  \\
2457112.747     &       0.27    &       -23.4763        &       0.0003  &       2457175.554     &       0.97    &       -34.8038        &       0.0009  \\
2457113.753     &       0.46    &       -13.2657        &       0.0003  &       2457175.721     &       0.01    &       -35.3598        &       0.0008  \\
2457137.747     &       0.93    &       -29.1695        &       0.0016  &       2457175.722     &       0.01    &       -35.3590        &       0.0008  \\
2457142.728     &       0.85    &       -5.3965 &       0.0012  &       2457176.545     &       0.16    &       -29.4283        &       0.0008  \\
2457143.703     &       0.04    &       -34.6591        &       0.0008  &       2457176.546     &       0.16    &       -29.4176        &       0.0008  \\
2457143.704     &       0.04    &       -34.6507        &       0.0007  &       2457176.723     &       0.19    &       -27.7502        &       0.0006  \\
2457144.709     &       0.22    &       -25.9233        &       0.0006  &       2457176.724     &       0.19    &       -27.7408        &       0.0006  \\
2457144.710     &       0.22    &       -25.9144        &       0.0006  &       2457177.526     &       0.34    &       -19.6200        &       0.0007  \\
2457145.712     &       0.41    &       -15.7395        &       0.0003  &       2457177.527     &       0.34    &       -19.6091        &       0.0007  \\
2457145.714     &       0.41    &       -15.7263        &       0.0003  &       2457177.598     &       0.36    &       -18.8642        &       0.0005  \\
2457146.695     &       0.59    &       -6.5524 &       0.0007  &       2457177.599     &       0.36    &       -18.8540        &       0.0006  \\
2457146.696     &       0.59    &       -6.5400 &       0.0007  &       2457178.543     &       0.53    &       -9.5137 &       0.0007  \\
2457147.726     &       0.79    &       3.0172  &       0.0012  &       2457178.544     &       0.53    &       -9.5015 &       0.0008  \\
2457147.727     &       0.79    &       3.0182  &       0.0012  &       2457178.718     &       0.56    &       -7.9788 &       0.0011  \\
2457148.704     &       0.97    &       -34.7264        &       0.0007  &       2457178.720     &       0.56    &       -7.9717 &       0.0020  \\
2457148.705     &       0.97    &       -34.7357        &       0.0008  &       2457204.523     &       0.37    &       -18.0216        &       0.0005  \\
2457153.719     &       0.90    &       -22.7805        &       0.0016  &       2457204.524     &       0.37    &       -18.0121        &       0.0005  \\
2457153.720     &       0.90    &       -22.8391        &       0.0017  &       2457205.547     &       0.56    &       -8.0510 &       0.0007  \\
2457154.659     &       0.08    &       -32.9507        &       0.0028  &       2457205.548     &       0.56    &       -8.0441 &       0.0007  \\
2457154.661     &       0.08    &       -32.9407        &       0.0027  &       2457206.455     &       0.73    &       0.8814  &       0.0008  \\
2457156.745     &       0.47    &       -12.6738        &       0.0016  &       2457206.457     &       0.73    &       0.8987  &       0.0007  \\
2457157.695     &       0.64    &       -4.2804 &       0.0025  &       2457206.538     &       0.75    &       1.7037  &       0.0009  \\
2457157.697     &       0.64    &       -4.2640 &       0.0023  &       2457206.539     &       0.75    &       1.7139  &       0.0009  \\
2457159.734     &       0.02    &       -35.0911        &       0.0010  &       2457206.723     &       0.78    &       2.8518  &       0.0013  \\
2457159.735     &       0.02    &       -35.0878        &       0.0010  &       2457206.724     &       0.78    &       2.8540  &       0.0015  \\
2457169.546     &       0.85    &       -4.5949 &       0.0012  &       2457207.546     &       0.94    &       -30.6013        &       0.0009  \\
2457169.547     &       0.85    &       -4.6461 &       0.0013  &       2457207.547     &       0.94    &       -30.6337        &       0.0009  \\
2457169.614     &       0.86    &       -8.7664 &       0.0012  &       2457208.471     &       0.11    &       -31.8710        &       0.0010  \\
2457169.615     &       0.87    &       -8.8388 &       0.0014  &       2457208.472     &       0.11    &       -31.8670        &       0.0010  \\
2457170.558     &       0.04    &       -34.5581        &       0.0010  &       2457208.621     &       0.14    &       -30.5734        &       0.0011  \\
2457170.559     &       0.04    &       -34.5525        &       0.0009  &       2457209.466     &       0.29    &       -22.3141        &       0.0008  \\
2457170.718     &       0.07    &       -33.4392        &       0.0011  &       2457209.467     &       0.29    &       -22.3028        &       0.0010  \\
2457170.719     &       0.07    &       -33.4328        &       0.0011  &       2457209.708     &       0.34    &       -19.8308        &       0.0006  \\
2457170.721     &       0.07    &       -33.4184        &       0.0009  &       2457209.709     &       0.34    &       -19.8196        &       0.0006  \\
2457170.721     &       0.07    &       -33.4142        &       0.0008  &       2457210.515     &       0.49    &       -11.7732        &       0.0009  \\
2457171.536     &       0.22    &       -25.9946        &       0.0010  &       2457210.515     &       0.49    &       -11.7647        &       0.0009  \\
2457171.540     &       0.22    &       -25.9575        &       0.0009  &       2457210.623     &       0.51    &       -10.7254        &       0.0008  \\
2457171.600     &       0.24    &       -25.3673        &       0.0006  &       2457210.624     &       0.51    &       -10.7162        &       0.0008  \\
2457171.602     &       0.24    &       -25.3431        &       0.0006  &       2457255.670     &       0.90    &       -22.2876        &       0.0014  \\
2457172.596     &       0.42    &       -15.1771        &       0.0005  &       2457255.671     &       0.90    &       -22.3447        &       0.0013  \\
2457172.598     &       0.42    &       -15.1607        &       0.0005  &       2457255.745     &       0.92    &       -26.4495        &       0.0010  \\
2457172.712     &       0.44    &       -14.0256        &       0.0005  &       2457255.746     &       0.92    &       -26.5002        &       0.0009  \\
2457172.713     &       0.44    &       -14.0132        &       0.0005  &       2457255.748     &       0.92    &       -26.6127        &       0.0010  \\
2457173.530     &       0.59    &       -6.5085 &       0.0010  &       2457255.749     &       0.92    &       -26.6636        &       0.0010  \\
2457173.532     &       0.60    &       -6.4944 &       0.0010  &       2457255.751     &       0.92    &       -26.7676        &       0.0010  \\
2457173.725     &       0.63    &       -4.8069 &       0.0007  &       2457255.752     &       0.92    &       -26.8161        &       0.0010  \\
2457173.726     &       0.63    &       -4.7941 &       0.0007  &       2457270.699     &       0.70    &       -0.9438 &       0.0008  \\
2457174.531     &       0.78    &       2.8942  &       0.0013  &       2457270.700     &       0.71    &       -0.9308 &       0.0008  \\
2457174.535     &       0.78    &       2.8976  &       0.0014  &       2457271.731     &       0.90    &       -20.0635        &       0.0007  \\
2457174.597     &       0.79    &       2.8737  &       0.0008  &               &               &               &               \\
\hline
days     &    &  \kms   &  \kms   & days     &    &  \kms   &  \kms  \\
\hline 
\end{tabular}
\end{center}
\end{table*}

\begin{table}
\begin{center}
\caption{Spectral lines used this study. 
\label{Tab_Lines}}
\begin{tabular}{rlcccc}
\hline \hline \noalign{\smallskip}

No. & El.   &   Wavelength (\AA)   & Ep (eV) & $\log(gf)$ \\ 
\hline
1 & \ion{Fe}{I} &       4683.560        &       2.831   &       -2.319  \\ 
2 & \ion{Fe}{I} &       4896.439        &       3.883   &       -2.050  \\ 
3 & \ion{Ni}{I} &       5082.339        &       3.658   &       -0.540  \\ 
4 & \ion{Fe}{I} &       5367.467        &       4.415   &       0.443   \\ 
5 & \ion{Fe}{I} &       5373.709        &       4.473   &       -0.860  \\ 
6 & \ion{Fe}{I} &       5383.369        &       4.312   &       0.645   \\ 
7 & \ion{Ti}{II}        &       5418.751        &       1.582   &       -2.110  \\ 
8 & \ion{Fe}{I} &       5576.089        &       3.430   &       -1.000  \\ 
9 & \ion{Fe}{I} &       5862.353        &       4.549   &       -0.058  \\ 
10 & \ion{Fe}{I}        &       6003.012        &       3.881   &       -1.120  \\ 
11 & \ion{Fe}{I}        &       6024.058        &       4.548   &       -0.120  \\ 
12 & \ion{Fe}{I}        &       6027.051        &       4.076   &       -1.089  \\ 
13 & \ion{Fe}{I}        &       6056.005        &       4.733   &       -0.460  \\ 
14 & \ion{Si}{I}        &       6155.134        &       5.619   &       -0.400  \\ 
15 & \ion{Fe}{I}        &       6252.555        &       2.404   &       -1.687  \\ 
16 & \ion{Fe}{I}        &       6265.134        &       2.176   &       -2.550  \\ 
17 & \ion{Fe}{I}        &       6336.824        &       3.686   &       -0.856  \\  

\hline \noalign{\smallskip}
\end{tabular}
\end{center}
\end{table}

\section{Inverse Baade-Wesselink projection factors derived from observations}\label{s_BWinv}

\subsection{Using the cross-correlated radial velocity curve}\label{ss_RVcc}

We describe the interferometric version of the  BW method as follows.  We apply a classical ${\chi}^{2} $ minimization 

\begin{equation}
\label{chi2sum} {\chi}^{2} = \sum_{i}{\frac{(\theta_{\rm
obs}(\phi_{i}) - \theta_{\rm model}(\phi_{i}))^2}{\sigma_{\rm
obs}(\phi_{i})^2}},
\end{equation}

where 

\begin{figure}[htbp]
\begin{center}
\resizebox{1.0\hsize}{!}{\includegraphics[clip=true]{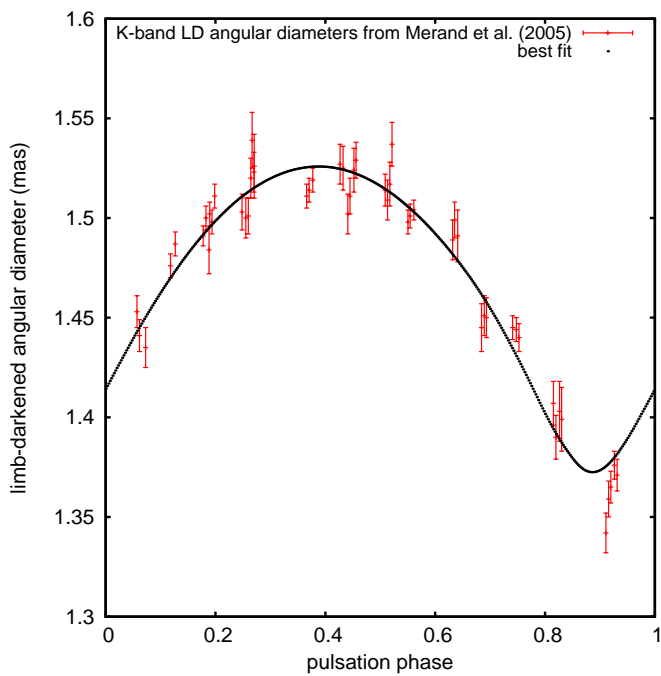}}
\end{center}
\caption{ Inverse BW method  applied to FLUOR/CHARA data \citep{merand05} considering a distance for $\delta$~Cep of $d=272 \pm 3 \pm  5$ pc \citep{majaess12} and the HARPS-N cross-correlated radial velocity curve. The black dotted line corresponds to the best fit.} \label{f1b_theta}
\end{figure}

\begin{itemize}
\item $ \theta_{\rm obs}(\phi_{i})$ are the interferometric limb-darkened angular diameters obtained from FLUOR/CHARA observations \citep{merand05}, with $\phi_{i}$ the pulsation phase corresponding to the $i$-th measurement (Fig.~\ref{f1b_theta}); 
\item $\sigma_{\rm obs}(\phi_{i})$ are the statistical uncertainties corresponding to FLUOR/CHARA measurements;
\item  $\theta_{\rm model}(\phi_{i})$ are the modeled limb-darkened angular diameters, defined as
\begin{equation}\label{diam_mod}
\theta_{\rm model}(\phi_{i}) = \overline{\theta} +
9.3009\,\frac{p_\mathrm{cc-g}}{d} \left(\int RV_\mathrm{cc-g}(\phi_{i})d\phi_{i}\right) [{\rm mas}],
\end{equation}
\end{itemize}
where the conversion factor 9.3009 is defined using the solar radius given in \citet{prsa16}.

The $RV_\mathrm{cc-g}(\phi_{i})$ is the interpolated HARPS-N cross-correlated radial velocity curve shown in panel a) of  Fig.~\ref{f1a_RVcc}. It is obtained  using 
the Gaussian fit, i.e., the most common approach in the literature. The parameters $\overline{\theta}$ and $p_\mathrm{cc-g}$ are the mean angular diameter of the star (in mas) and the projection factor (associated with  the Gaussian fit of the CCFs), respectively, while $d$ is the distance to the star. The quantities $\overline{\theta}$ and $p_\mathrm{cc-g}$ are fit in order to minimize ${\chi}^{2}$, while $d$ is fixed to $d=272 \pm 3 (stat.) \pm  5 (syst.)$ pc \citep{majaess12}. 

\begin{figure}[htbp]
\begin{center}
\resizebox{0.9\hsize}{!}{\includegraphics[clip=true]{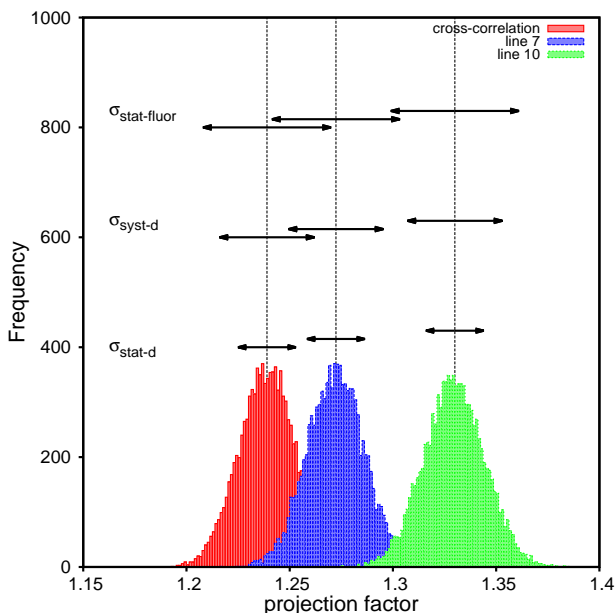}}
\end{center}
\caption{Inverse BW method described in Sect.~\ref{s_BWinv}  applied considering a Gaussian distribution for the distance. The resulting projection factor distributions are shown in three particular cases: when using the cross-correlated radial velocity curve RV$_\mathrm{cc-g}$ (red), the RV$_\mathrm{c}$ radial velocity curve of line 7 (blue), and the RV$_\mathrm{c}$ radial velocity curve of line 10 (green). The quantities $\sigma_\mathrm{stat-d}$ and $\sigma_\mathrm{syst-d}$ are the uncertainties on the projection factor due to the statistical and systematical uncertainties on the distance, respectively. The quantity $\sigma_\mathrm{stat-fluor}$ stems from  the statistical uncertainties on the FLUOR/CHARA interferometric measurements.} \label{Fig_mcpinv}
\end{figure}

We find $\overline{\theta}=1.466\pm0.007$ mas and $p_\mathrm{cc-g}=1.239 \pm 0.031$, where the uncertainty on the projection factor (hereafter $\sigma_\mathrm{stat-fluor}$) is  about 2.5\% and stems from FLUOR/CHARA angular diameter measurements. If we apply this  procedure 10,000 times  using a Gaussian distribution for the assumed distance that is centered at $272$~pc and has a half width at half maximum (HWHM) of $3$~pc (corresponding to the statistical precision on the distance of \citealt{majaess12}), then we obtain a symmetric distribution for the 10,000 values of $p_\mathrm{cc-g}$ with a HWHM $\sigma_\mathrm{stat-d}=0.014$.  If the distance is set to $277$~pc and $267$~pc (corresponding to the systematical uncertainty of $\pm5$~pc of \citealt{majaess12}), we find $p=1.262$ and $p=1.216$, respectively. We have a systematical uncertainty $\sigma_\mathrm{syst-d}=0.023$ for the projection factor. We thus find $p_\mathrm{cc-g}=1.239\pm0.031$~($\sigma_\mathrm{stat-fluor}$) $\pm0.014$~($\sigma_\mathrm{stat-d}$) $\pm0.023$~($\sigma_\mathrm{syst-d}$).  These results are illustrated in Fig.~\ref{Fig_mcpinv}. Interestingly, if we use the RV$_\mathrm{cc-c}$ curve in Eq.~\ref{diam_mod} (still keeping $d=272$~pc), we obtain $p_\mathrm{cc-c}=1.272$ while the uncertainties remain unchanged.

The distance of $\delta$~Cep obtained by \citet{majaess12} is an average of {\it Hipparcos} \citep{leeuw2007b} and HST \citep{benedict02} trigonometric parallaxes, together with the cluster main sequence fitting distance. If we rely on the direct distance to $\delta$ Cep obtained only by HST  \citep{benedict02}, i.e., $273\pm11$~pc, then $\sigma_\mathrm{stat-d}$ is larger with a value of $0.050$ (compared to 0.014 when relying on \citealt{majaess12}). 
A distance $d=244~\pm~10$~pc  was obtained by  \citet{anderson15a} from the reanalysis of the {\it Hipparcos} astrometry of $\delta$~Cep. It leads to a very small projection factor $p_\mathrm{cc-g}=1.14\pm0.031\pm0.047$ compared to other values listed in Table~\ref{tab_history}.


 
\subsection{Using the first moment radial velocity curves corresponding to individual spectral lines}\label{ss_RVci}

\begin{figure*}[htbp]
\begin{center}
\resizebox{0.8\hsize}{!}{\includegraphics[clip=true]{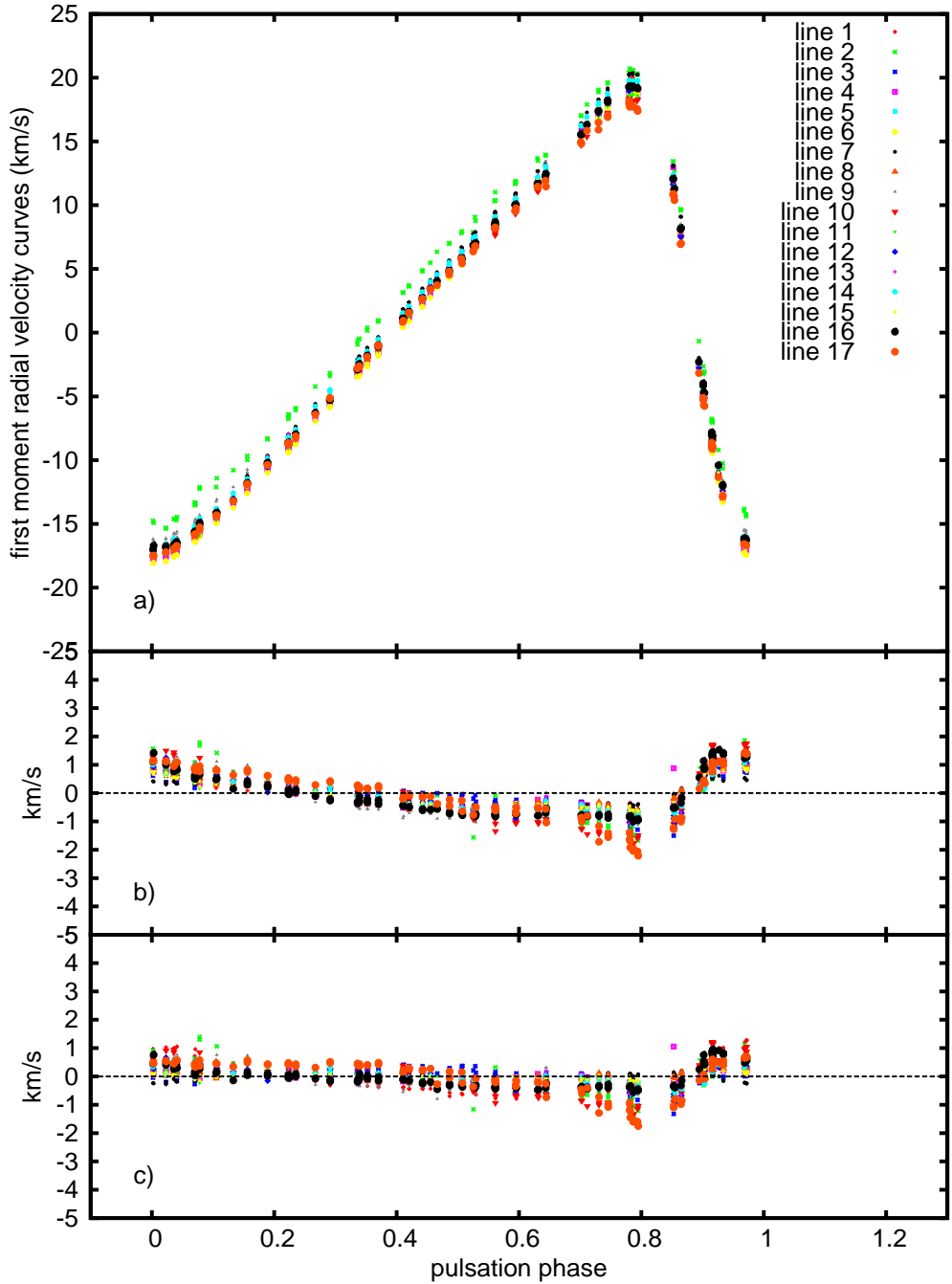}}
\end{center}
\vspace{-1cm}
\caption{ (a) First moment radial velocity curves ($RV_\mathrm{c}$) associated with the 17 lines of Table~\ref{Tab_Lines}. The $\gamma$-velocity associated with the cross-correlated radial velocity curve, RV$_\mathrm{cc-g}$ in Fig.~\ref{f1a_RVcc}a,  has been removed from these curves, i.e., $V_\mathrm{\gamma}= -16.95$ \kms. The $\gamma$ offset residuals are known to be related to the k-term \citep{nardetto08a}. The residuals between the $RV_\mathrm{c}$ curves and the RV$_\mathrm{cc-g}$ in Fig. ~\ref{f1a_RVcc}a (resp.  RV$_\mathrm{cc-c}$)  are plotted in panel (b) (resp. (c)) after correcting each $RV_\mathrm{c}$ curve from its $V_\mathrm{\gamma}$ value.} \label{f1c_RVci}
\end{figure*}

We use the 17 unblended spectral lines (Table~\ref{Tab_Lines}) that were previously selected for an analysis of  eight Cepheids with periods ranging from 4.7 to 42.9~d  \citep{nardetto07}. These lines remain unblended for every  pulsation phase of the Cepheids considered. Moreover, they
were carefully selected in order to represent a wide range of depths, hence to measure the atmospheric velocity gradient
(Sect.~\ref{ss_grad}). For each of these lines, we derive the centroid velocity
($RV_{\mathrm c}$), i.e., the first moment of the spectral line profile, estimated as 

\begin{equation} \label{Eq_CDG}
 RV_{\mathrm c} = \frac{\int_{\rm line} \lambda S(\lambda) d\lambda}{\int_{\rm line} S(\lambda) d\lambda}
,\end{equation}where $S(\lambda)$ is the observed line profile. The radial velocity measurements associated with the spectral lines are presented in Fig.~\ref{f1c_RVci}a together with the interpolated RV$_\mathrm{cc-g}$ curve. The RV$_\mathrm{c}$ curves plotted have been corrected for the $\gamma$-velocity value corresponding to the RV$_\mathrm{cc-g}$ curve, i.e., $V_\mathrm{\gamma}=-16.95$~\kms. The residuals, i.e., the $\gamma$-velocity offsets, between the curves of Fig.~\ref{f1c_RVci}a are related to the line asymmetry and the k-term value (see \citealp{nardetto08a} for Cepheids and \citealp{nardetto13, nardetto14} for other types of pulsating stars). This will be analyzed in a forthcoming paper. The final RV$_\mathrm{c}$ curves used in the inverse BW approach are corrected from their own residual $\gamma$-velocity in such a way that the interpolated curve has an average of zero. The residual of these curves compared to the RV$_\mathrm{cc-g}$ and RV$_\mathrm{cc-c}$ curves are shown in panel b) and c), respectively. We then apply the same method as in Sect.~\ref{ss_RVcc}. The measured projection factor values $p_\mathrm{obs}$($k$) associated with each spectral line $k$ are listed in Table~\ref{Tab_pf}. The statistical and systematical uncertainties in the case of individual lines are the same as those found when using the RV$_\mathrm{cc-g}$ curve. The projection factor values range from $1.273$ (line 7) to $1.329$ (line 10), whereas the value corresponding to the cross-correlation method is $1.239$ (see Fig.~\ref{Fig_mcpinv}). This shows that the projection factor depends significantly on the method used to derive the radial velocity and the spectral line considered. To analyze these values it is possible to  use hydrodynamical simulations and the projection factor decomposition into three terms introduced by \citet{nardetto07}. 

\begin{table*}
\begin{center}
\caption{The observational quantities, $D$, $p_\mathrm{obs}$, and $f_\mathrm{grad}$, are listed for each line of Table~\ref{Tab_Lines}.  The quantities $p_\mathrm{hydro}$, $p_\mathrm{0}$ and $f_\mathrm{o-g}$ are derived from hydrodynamical model.} \label{Tab_pf}
\begin{tabular}{|c|ccc||cc|}
\hline \hline \noalign{\smallskip} 
   & \multicolumn{3}{c||}{HARPS-N observations} & \multicolumn{2}{c|}{Hydrodynamical model}  \\

        line    &       D$^{\mathrm{(a)}}$                               &       $p_\mathrm{obs \pm \sigma_\mathrm{fluor-stat} \pm \sigma_\mathrm{d-stat} \pm \sigma_\mathrm{d-syst}}$$^{\mathrm{(b)}}$                            &       $f_\mathrm{grad}$$^{\mathrm{(c)}}$                               &       $p_\mathrm{hydro}$$^{\mathrm{(d)}}$                              &       $p_\mathrm{0}$$^{\mathrm{(e)}}$  \\ 
        \hline
                \hline
        line 1  &       0.281   $_\mathrm{ \pm  0.001   }$      &       1.287   $_\mathrm{ \pm     0.031 \pm       0.014 \pm 0.023 }$      &       0.983   $_\mathrm{ \pm     0.007   }$      &       1.307           &       1.360   \\ 
        line 2  &       0.147   $_\mathrm{ \pm  0.001   }$      &       1.323   $_\mathrm{ \pm     0.031  \pm      0.014 \pm 0.023 }$      &       0.991   $_\mathrm{ \pm     0.004   }$      &       1.328           &       1.365   \\ 
        line 3  &       0.348   $_\mathrm{ \pm  0.001   }$      &       1.280   $_\mathrm{ \pm     0.031  \pm      0.014 \pm 0.023 }$      &       0.979   $_\mathrm{ \pm     0.009   }$      &       1.304           &       1.369   \\ 
        line 4  &       0.573   $_\mathrm{ \pm  0.001   }$      &       1.282   $_\mathrm{ \pm     0.031  \pm      0.014 \pm 0.023 }$      &       0.967   $_\mathrm{ \pm     0.014   }$      &       1.292           &       1.375   \\ 
        line 5  &       0.297   $_\mathrm{ \pm  0.001   }$      &       1.297   $_\mathrm{ \pm     0.031  \pm      0.014 \pm 0.023 }$      &       0.982   $_\mathrm{ \pm     0.008   }$      &       1.309           &       1.375   \\ 
        line 6  &       0.612   $_\mathrm{ \pm  0.001   }$      &       1.282   $_\mathrm{ \pm     0.031  \pm      0.014 \pm 0.023 }$      &       0.964   $_\mathrm{ \pm     0.015   }$      &       1.288           &       1.375   \\ 
        line 7  &       0.562   $_\mathrm{ \pm  0.001   }$      &       1.273   $_\mathrm{ \pm     0.031  \pm      0.014 \pm 0.023 }$      &       0.967   $_\mathrm{ \pm     0.014   }$      &       1.272           &       1.376   \\ 
        line 8  &       0.500   $_\mathrm{ \pm  0.001   }$      &       1.285   $_\mathrm{ \pm     0.031  \pm      0.014 \pm 0.023 }$      &       0.971   $_\mathrm{ \pm     0.012   }$      &       1.297           &       1.378   \\ 
        line 9  &       0.364   $_\mathrm{ \pm  0.001   }$      &       1.307   $_\mathrm{ \pm     0.031  \pm      0.014 \pm 0.023 }$      &       0.979   $_\mathrm{ \pm     0.009   }$      &       1.302           &       1.383   \\ 
        line 10 &       0.326   $_\mathrm{ \pm  0.001   }$      &       1.329   $_\mathrm{ \pm     0.031  \pm      0.014 \pm 0.023 }$      &       0.981   $_\mathrm{ \pm     0.008   }$      &       1.301           &       1.387   \\ 
        line 11 &       0.429   $_\mathrm{ \pm  0.001   }$      &       1.295   $_\mathrm{ \pm     0.031  \pm      0.014 \pm 0.023 }$      &       0.975   $_\mathrm{ \pm     0.011   }$      &       1.304           &       1.385   \\ 
        line 12 &       0.283   $_\mathrm{ \pm  0.001   }$      &       1.300   $_\mathrm{ \pm     0.031  \pm      0.014 \pm 0.023 }$      &       0.983   $_\mathrm{ \pm     0.007   }$      &       1.312           &       1.385   \\ 
        line 13 &       0.290   $_\mathrm{ \pm  0.002   }$      &       1.294   $_\mathrm{ \pm     0.031  \pm      0.014 \pm 0.023 }$      &       0.983   $_\mathrm{ \pm     0.008   }$      &       1.304           &       1.386   \\ 
        line 14 &       0.317   $_\mathrm{ \pm  0.002   }$      &       1.292   $_\mathrm{ \pm     0.031  \pm      0.014 \pm 0.023 }$      &       0.981   $_\mathrm{ \pm     0.008   }$      &       1.310           &       1.387   \\ 
        line 15 &       0.497   $_\mathrm{ \pm  0.001   }$      &       1.280   $_\mathrm{ \pm     0.031  \pm      0.014 \pm 0.023 }$      &       0.971   $_\mathrm{ \pm     0.012   }$      &       1.290           &       1.389   \\ 
        line 16 &       0.348   $_\mathrm{ \pm  0.001   }$      &       1.301   $_\mathrm{ \pm     0.031  \pm      0.014 \pm 0.023 }$      &       0.979   $_\mathrm{ \pm     0.009   }$      &       1.301           &       1.389   \\ 
        line 17 &       0.366   $_\mathrm{ \pm  0.001   }$      &       1.323   $_\mathrm{ \pm     0.031  \pm      0.014 \pm 0.023 }$      &       0.978   $_\mathrm{ \pm     0.009   }$      &       1.297           &       1.390   \\ 

\hline \noalign{\smallskip}
\end{tabular}
\end{center}
\begin{list}{}{}
\item[$^{\mathrm{(a)}}$]  The line depth $D$ is calculated at minimum radius of the star. 
\item[$^{\mathrm{(b)}}$]  The observational projection factors $p_\mathrm{obs}$ is derived from HARPS-N and FLUOR/CHARA interferometric data using the inverse BW approach (Sect.~\ref{ss_RVci}). 
\item[$^{\mathrm{(c)}}$]  The $f_\mathrm{grad}$ coefficient involved in the projection factor decomposition ($p =  p_\mathrm{0} f_\mathrm{grad} f_\mathrm{o-g}$, \citealt{nardetto07}) is derived from Eq.~\ref{Eq_grad2} and  Eq.~\ref{Eq_grad_O} (Sect.~\ref{ss_grad}). 
\item[$^{\mathrm{(d)}}$]   The inverse projection factors $p_\mathrm{hydro}$ is calculated with the hydrodynamical model (Sect.~\ref{ss_BWinv_model}) 
\item[$^{\mathrm{(e)}}$] The modeled geometric projection factor $p_\mathrm{0}$ is derived in the continuum next to each spectral line. The $f_\mathrm{o-g}$ quantity, which is obtained using the projection factor decomposition ($f_\mathrm{o-g}= \frac{p_\mathrm{obs}}{p_{\mathrm{o}}\,f_{\mathrm{grad}}}$), is derived in Sect.~\ref{s_fog}.
\end{list}
\end{table*}

\section{Comparing the hydrodynamical model of $\delta$~Cep with observations}\label{s_hydro}

Our best model of $\delta$~Cep was presented in \citet{nardetto04} and is computed using the code by \citet{fokin91}. 
The hydrodynamical model requires only five input fundamental parameters: $M=4.8\,\Mo$, $L=1995\,\Lo$, $\te=5877$~K, $Y=0.28$, and $Z=0.02$. 
At the limit cycle the pulsation period is $5.419$~d, very close (1\%) to the observed value.

\subsection{ Baade-Wesselink projection factors derived from the hydrodynamical model}\label{ss_BWinv_model}

\begin{figure}[htbp]
\begin{center}
\resizebox{1.0\hsize}{!}{\includegraphics[clip=true]{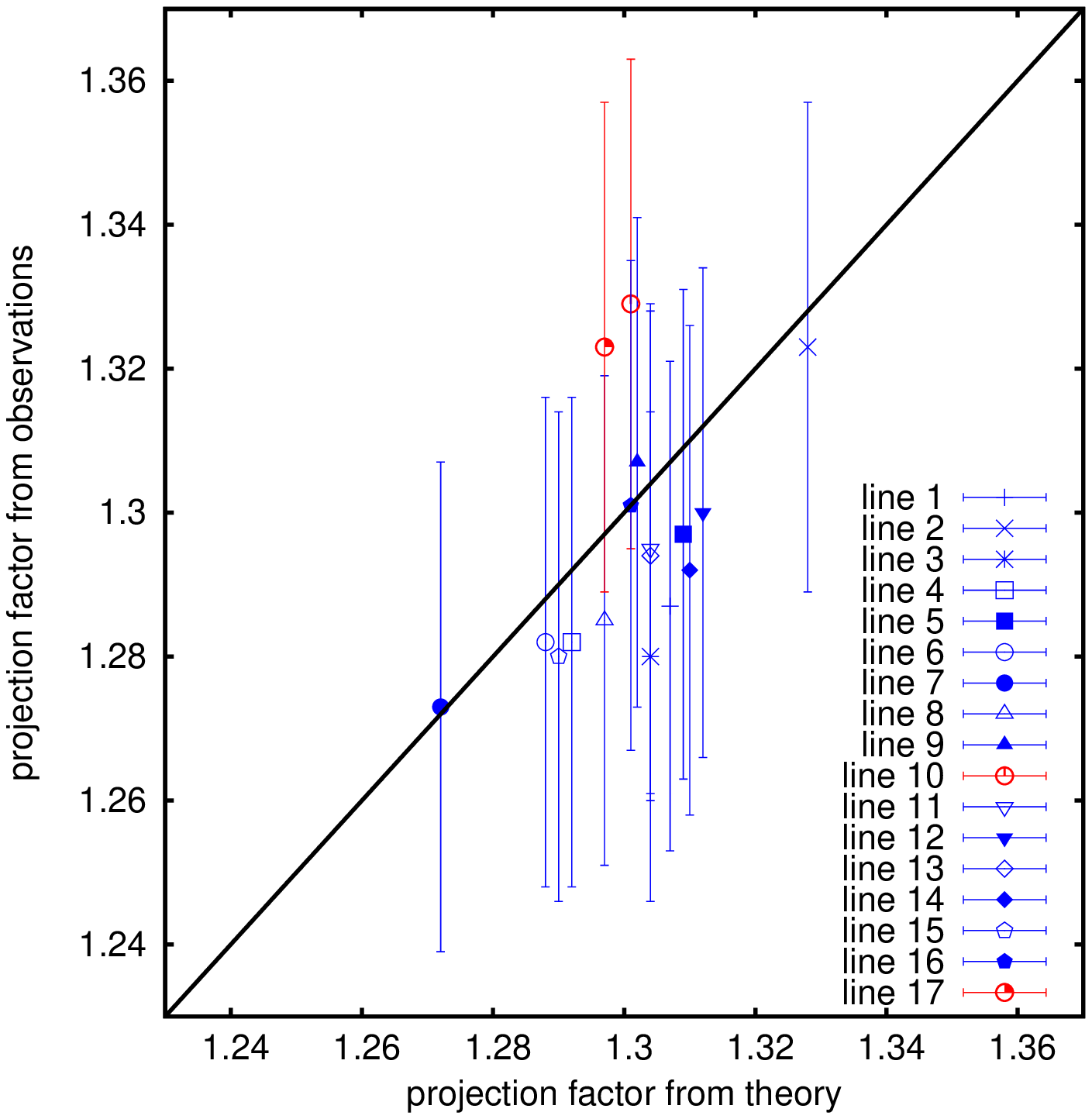}}
\end{center}
\caption{Observational projection factors derived using the inverse BW approach described in Sect.~\ref{s_BWinv}  compared to the theoretical values obtained from the hydrodynamical model (Sect.~\ref{ss_BWinv_model}).} \label{f2a_pf}
\end{figure}

The projection factors are derived directly from the hydrodynamical code for each spectral line of Table~\ref{Tab_Lines} following the definition and procedure described in \citet{nardetto07}. The computed projection factors  range from $1.272$ (line 7) to $1.328$ (line 2). In Fig.~\ref{f2a_pf} we plot the theoretical projection factors ($p_\mathrm{hydro}$) as a function of the observational values (listed in Table~\ref{Tab_pf}). In this figure the statistical uncertainties on the observational projection factors, i.e., $\sigma_\mathrm{stat-fluor}$ and $\sigma_\mathrm{stat-d}$, have been summed quadratically (i.e., $\sigma=0.034$). The agreement is excellent 
since the most discrepant lines (10 and 17, in red in the figure) show observational projection factor values only about 1$\sigma$ larger than the theoretical values. 


\subsection{Atmospheric velocity gradient of $\delta$~Cep}\label{ss_grad}

In  \citet{nardetto07}, we split the projection factor into three quantities: $p= p_{\mathrm{o}}\,f_{\mathrm{grad}}\,f_{\mathrm{o-g}}$, where $p_\mathrm{0}$ is the geometrical projection factor (linked to the limb darkening of the star); $f_\mathrm{grad}$, which  is a cycle-integrated quantity linked to the velocity gradient in the atmosphere of the star (i.e., between the considered line-forming region and the photosphere); and $f_\mathrm{o-g}$, which is the relative motion of the optical pulsating photosphere with respect to the corresponding mass elements.



We derive $f_{\mathrm{grad}}$ directly from HARPS-N observations. In \citet{nardetto07}, we  showed that the line depth taken at the minimum radius phase (hereafter $D$) traces the height of the line-forming regions in such a way that the projection factor decomposition is possible. By comparing $ \Delta RV_{\mathrm{c}}$ with the depth, $D$, of the 17 selected spectral lines listed in Table~\ref{Tab_Lines}, we  directly measure $f_\mathrm{grad}$. If we define $a_0$ and $b_0$ as the slope and zero-point of the linear correlation (the photosphere being the zero line depth),

\begin{equation} \label{Eq_grad}
\Delta RV_{\mathrm{c}}= a_0 D + b_0, 
\end{equation}then the velocity gradient correction on the projection factor is

\begin{equation} \label{Eq_grad2}
 f_{\mathrm{grad}}= \frac{b_0}{a_0D+ b_0}. 
\end{equation}

In \citet{anderson16b} the atmospheric velocity gradient is defined as the difference between velocities determined using weak and strong lines at each pulsation phase. In our description of the projection factor, $f_\mathrm{grad}$ is calculated with Eqs.~\ref{Eq_grad} and ~\ref{Eq_grad2}, i.e., considering the amplitude of the radial velocity curves from individual lines. In the following we thus refer to $f_\mathrm{grad}$ as a cycle integrated quantity. 

\begin{figure*}[htbp]
\begin{center}
\resizebox{0.8\hsize}{!}{\includegraphics[clip=true]{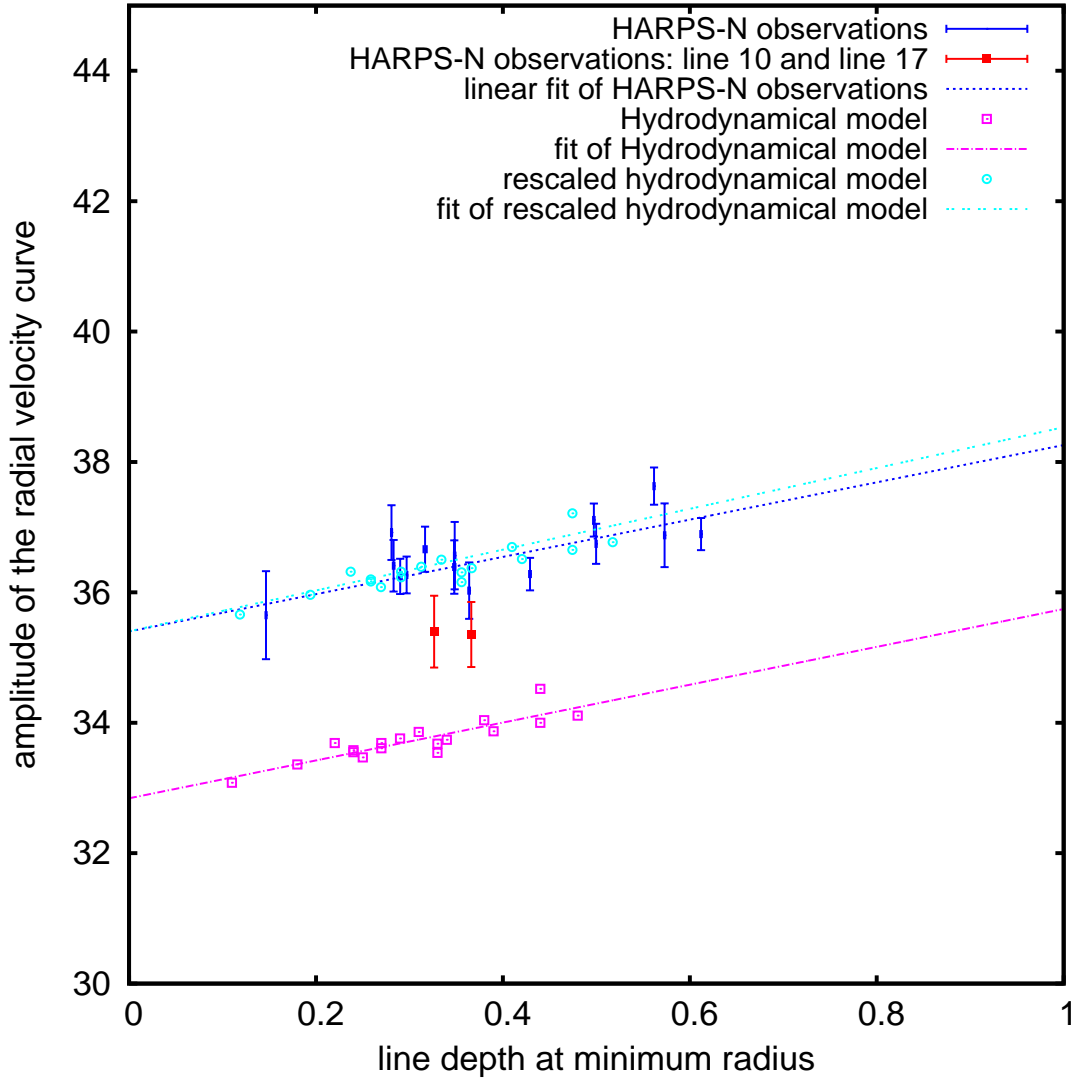}}
\end{center}
\caption{Amplitude of the radial velocity curves for the 17 spectral lines listed in Table~\ref{Tab_Lines}  plotted versus the line depth for the hydrodynamical model (magenta squares), and for the HARPS-N spectroscopic observations (blue dots, except lines 10 and 17 plotted with filled red squares). A rescale of the model with a multiplying factor (i.e., $f_\mathrm{c}$) is necessary to fit the data (light blue circles). 
} \label{Fig_grad}
\end{figure*}

In Fig.~\ref{Fig_grad} we plot the HARPS-N measurements with blue dots, except for lines 10 and 17 for which we use red squares. These two values are about 2$\sigma$ below the other measurements, as already noted  in Fig.~\ref{f2a_pf}. After fitting all measured $\Delta RV_{\mathrm{c}}\mathrm{[obs]}$ with a linear relation we find

\begin{equation} \label{Eq_grad_O}
\Delta RV_{\mathrm{c}}\mathrm{[obs]}= [2.86 \pm 0.84] D +  [35.40 \pm 0.36] 
.\end{equation}

The reduced $\chi^2$ is 1.4 and decreases to 1.1 if lines 10 and 17 are not considered (but with approximately the same values of $a_0$ and $b_0$). For comparison, the reduced $\chi^2$ is 2.5 if a horizontal line is fitted. 
The same quantities derived from the hydrodynamical model are shown in Fig.~\ref{Fig_grad} ($\Delta RV_{\mathrm{c}}\mathrm{[mod]}$, magenta squares). The corresponding relation is
 \begin{equation} \label{Eq_grad_M}
\Delta RV_{\mathrm{c}}\mathrm{[mod]}= [2.90 \pm 0.39] D +  [32.84 \pm 0.13] 
.\end{equation}

The slopes of Eq.~\ref{Eq_grad_O} and \ref{Eq_grad_M} are consistent, while the theoretical zero-point is about 2.6 \kms below the corresponding observational value, which means that the amplitudes of the theoretical radial velocity curves are 2.6 \kms (or 7.8\%) smaller on average. Such disagreement occurs because our code is self-consistent, i.e., the radial velocity curve is not used as an input like in a {\it piston} code, and because the treatment of convection in the code is missing which can slightly bias (by a few percent) the input fundamental parameters. The two- or three-dimensional models that properly describe the coupling between the pulsation and the convection \citep{geroux15, houdek15} are currently not providing synthetic profiles, hence preventing the calculation of the projection factor. Therefore, we rely on our purely radiative hydrodynamical code (as previously done in \citealp{nardetto04, nardetto07}) to study the atmosphere of Cepheids.  Its consistency with the spectroscopic and interferometric observables is satisfactory as soon as we consider a multiplying correcting factor of $f_\mathrm{c}=1.078$. Consequently, Eq.~\ref{Eq_grad_M} becomes

\begin{equation} \label{Eq_grad_M_res}
\Delta RV_{\mathrm{c}}\mathrm{[mod]}= [2.90 D + 32.84] *f_\mathrm{c}.
\end{equation}

The corresponding values are shown in Fig.~\ref{Fig_grad} with light blue squares, and the agreement with observations is now excellent. If the theoretical amplitudes of the radial velocity curves are underestimated, why do we obtain the correct values of the projection factors in Sect.~\ref{ss_BWinv_model}? The answer is that the projection factor depends only on the ratio of pulsation to radial velocities. If the pulsation velocity curve has an amplitude that is  7.8\% larger, then the radial velocity curve (whatever the line considered) and the radius variation (see Sect.~\ref{ss_RV}), also have   amplitudes that are  7.8\% larger and the derived projection factor remains the same. 

Small differences in the velocity amplitudes between the RV$_\mathrm{cc-c}$ and RV$_\mathrm{c}$ curves (Fig.~\ref{f1c_RVci}c) are due
to the use of different methods and line samples (a full mask and 17 selected lines,
respectively).

\subsection{Angular diameter curve.}\label{ss_RV}

\begin{figure}[htbp]
\begin{center}
\resizebox{0.9\hsize}{!}{\includegraphics[clip=true]{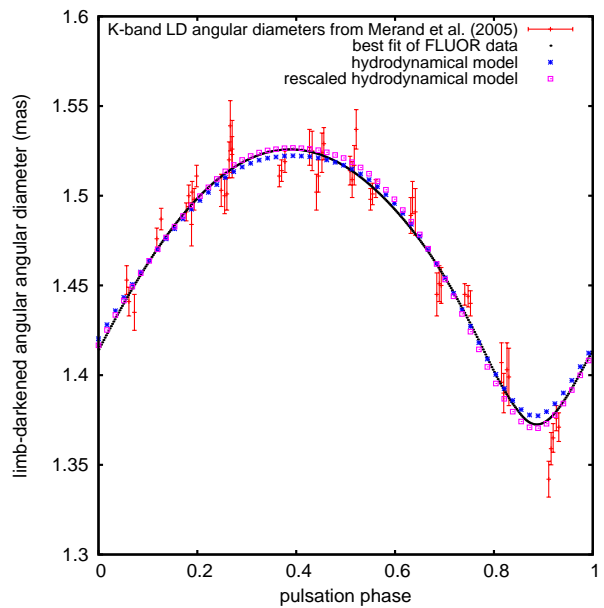}}
\end{center}
\caption{ FLUOR/CHARA limb-darkened angular diameter curve  compared to the hydrodynamical model (see the text).} \label{fig_ang} 
\end{figure}


In Fig. \ref{fig_ang}, we compare our best-fit infrared angular diameters from FLUOR/CHARA (same curve as Fig.~\ref{f1b_theta}) with the photospheric angular diameters derived directly from the model assuming a distance of $d=272~$pc \citep{majaess12}. Following the projection factor decomposition, this photospheric angular diameter 
is calculated by integrating the pulsation velocity associated with the photosphere of the star. 
We consider this to be the layer of the star for which the optical depth in the continuum (in the vicinity of the \ion{Fe}{I} 6003.012\,\AA\, spectral line) is $\tau_c=\frac{2}{3}$. However, to superimpose the computed photospheric angular diameter curve on the interferometric one, we  again need a correction factor $f_\mathrm{c}$. We find that the rescaled model (magenta open squares) is consistent with the solid line, which corresponds to the integration of the HARPS-N RV$_\mathrm{cc-g}$ curve (multiplied consistently by p$_\mathrm{cc-g}$). 

We rescaled the outputs of the hydrodynamical code, i.e., the atmospheric velocity gradient (Sect.~\ref{ss_grad}), the radial velocity, and the angular diameter curves, by the same quantity $f_\mathrm{c}$ in order to reproduce the observations satisfactorily. This scaling leaves the hydrodynamical projection factors unchanged and in agreement with the observational values (Table~\ref{Tab_Lines}). This can be seen using Eq.~\ref{diam_mod}: if we multiply each part of this equation by $f_\mathrm{c}$, the result in terms of the projection factor is unchanged.

\section{Determining $f_\mathrm{o-g}$}\label{s_fog}

\begin{figure}[htbp]
\begin{center}
\resizebox{1.0\hsize}{!}{\includegraphics[clip=true]{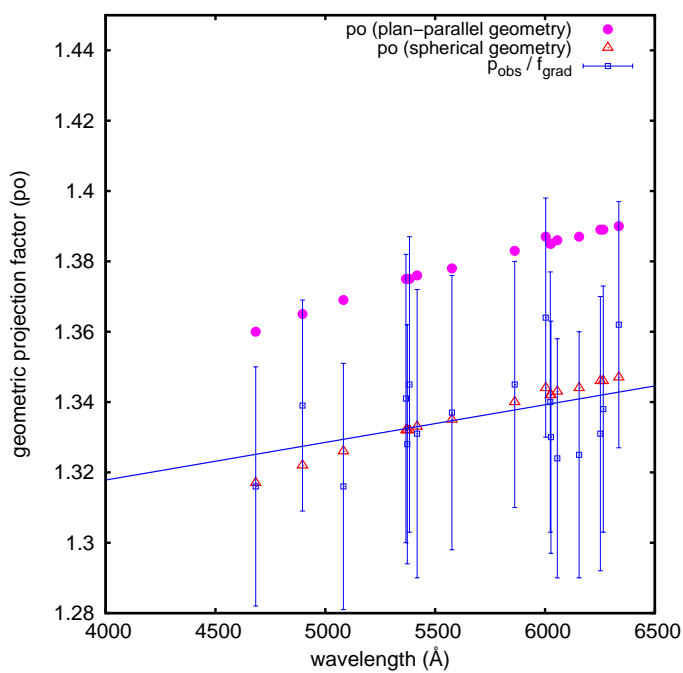}}
\end{center}
\caption{ Geometric projection factors calculated using radiative transfer in plane-parallel and in spherical geometry  used together with the observational quantity $\frac{p_\mathrm{obs}}{f_\mathrm{grad}}$ in order to derive $f_\mathrm{o-g}$ (see Sect.~\ref{s_fog}).} \label{Fig_po}
\end{figure}

The variable $f_{\mathrm{o-g}}$ is linked to the distinction between the {\it optical} and {\it gas} photospheric layers. The {\it optical} layer is the location where the continuum is  generated ($\tau_c=\frac{2}{3}$). 
The {\it gas} layer is the location of some mass element in the hydrodynamic model mesh where, at some moment in time, the photosphere is located. Given that the location of the photosphere moves through different mass elements as the star pulsates, the two layers have different velocities, hence it is necessary to define $f_{\mathrm{o-g}}$ in the projection factor decomposition. The $f_\mathrm{o-g}$ quantity is independent of the spectral line considered and is given by

\begin{equation} \label{Eq_pf_decomposition}
f_{\mathrm{o-g}} = \frac{p_\mathrm{obs}(k)}{p_{\mathrm{o}}(k)\,f_{\mathrm{grad}}(k)}\ ,
\end{equation}
where $k$ indicates the spectral line considered. From the previous sections, we now have the ability to derive $ f_{\mathrm{o-g}}$. In Sect.~\ref{s_BWinv}, we derived the projection factors $p_\mathrm{obs}(k)$ for 17 individual lines. In Sect.~\ref{ss_grad}, we  determined $f_{\mathrm{grad}}(k)= \frac{b_0}{a_0D_k+ b_0}$ using Eq.~\ref{Eq_grad2} (see Table~\ref{Tab_pf}). The last quantity required to derive $f_{\mathrm{o-g}}$ is the geometric projection factor  $p_\mathrm{o}$ (see Eq.~\ref{Eq_pf_decomposition}). There is currently no direct estimation of $p_\mathrm{o}$ for $\delta$~Cep. If we rely on the hydrodynamical model, $p_\mathrm{o}$ can be inferred from the intensity distribution next to the continuum of each spectral line. The list of geometric projection factors are listed in Table~\ref{Tab_pf} and plotted in Fig.~\ref{Fig_po} with magenta dots. These calculations are done in the plane-parallel radiative transfer approximation. On the other hand, \citet{neilson12}  showed that the p-factor differs significantly as a function of geometry where those from plane-parallel model atmospheres are 3-7\% greater then those derived from spherically symmetric models. Using their Table~1, we find for $\delta$~Cep a spherically symmetric geometrical projection factor of 1.342 in the R-band (i.e., with an effective wavelength of 6000~\AA). In Fig.~\ref{Fig_po}, if we shift our results (magenta dots) by $0.043$ in order to get 1.342 at 6000~\AA\ (a decrease of 3.2\%), we roughly estimate the geometrical projection factors in spherical geometry as a function of the wavelength (red open triangles). In Fig.~\ref{Fig_po}, we now plot the $\frac{p_\mathrm{obs}(k)}{f_{\mathrm{grad}}(k)}$ quantity for each individual spectral line 
with their corresponding uncertainties (blue open squares). If we divide the $\frac{p_\mathrm{obs}(k)}{f_{\mathrm{grad}}(k)}$ quantities obtained for each individual spectral line by the corresponding value of $p_{\mathrm{o}}(k)$ calculated in plane-parallel geometry, we obtain $f_{\mathrm{o-g}} = 0.975 \pm 0.002$, with a reduced $\chi^2$ of 0.13. This indicates that our uncertainties (the quadratic sum of $\sigma_\mathrm{stat-fluor}$, $\sigma_\mathrm{stat-d}$ and the statistical uncertainty on $f_\mathrm{grad}$) are probably overestimated. This value is several $\sigma$ greater than that found directly with the hydrodynamical model of $\delta$~Cep: $f_\mathrm{o-g}=0.963 \pm 0.005$ (\citealp{nardetto07}, their Table~5). Using the values of  $p_{\mathrm{o}}$ from \citet{neilson12} (blue open triangles), we obtain $f_{\mathrm{o-g}} = 1.006 \pm 0.002$. 
Thus, $f_{\mathrm{o-g}}$ depends significantly on the model used to calculate $p_{\mathrm{o}}$. 


\section{Conclusion}\label{s_conclusion}

Our rescaled hydrodynamical model of $\delta$~Cep is consistent with both spectroscopic and interferometric data modulo a rescaling factor that depends on the input parameters of the model ($M, L, T_\mathrm{eff}, Z$). In particular, it reproduces the observed amplitudes of the radial velocity curves associated with a selection of 17 unblended spectral lines as a function of the line depth in a very satisfactory way. This is a critical step for deriving the correct value of the projection factor. This strongly suggests that our decomposition of the projection factor into three physical terms is adequate. The next difficult step will be to measure  $p_\mathrm{0}$ directly from the next generation of visible-wavelength interferometers. With such values in hand, it will be possible to  derive $f_\mathrm{o-g}$ directly from observations.  

The projection factor is a complex quantity that is particularly sensitive to the definition of the radial velocity measurement. More details should be given  and perhaps a standard procedure applied in future analyses. For instance, the CCFs are generally fitted with a Gaussian. This produces a velocity value sensitive to  stellar rotation and to both the line width and depth  \citep{nardetto06a}. Thus,  additional biases in the distance determination are introduced, in particular when comparing  the projection factors of different  Cepheids. Conversely, investigating the projection factor of individual lines is useful for learning how to mitigate the impact of the radial velocity modulation \citep{anderson14} and the possible angular-diameter modulation \citep{anderson16} on the BW distances.

In this study, we found $p_\mathrm{cc-g}=1.24\pm0.04$ a value that is consistent with the \emph{Pp} relation of \citet{nardetto09}, i.e., $p_\mathrm{cc-g}=1.25\pm 0.05$. However, a disagreement is still found between the interferometric and the infrared surface-brightness approaches (and  their respective projection factors)  in the case of $\delta$~Cep \citep{ngeow12}, while an agreement is found for the long-period Cepheid $\ell$~Car \citep{kervella04d}. This suggests that the p-factors adopted in these approaches might be affected by something not directly related to the projection factor itself but rather to other effects in the atmosphere or close to it, such as a static circumstellar environment \citep{nardetto16a}. However, it is worth mentioning that the two methods currently provide {\it absolutely the same results in terms of distances}. After  a long history  (since \citealt{getting34}), the BW projection factor remains a key quantity in the calibration of the cosmic distance scale, and  one century after the discovery of the Period-Luminosity relation \citep{leavitt08}, Cepheid pulsation is still a distinct challenge.  With Gaia and other high-quality spectroscopic data, it will soon be  possible to better constrain the \emph{Pp} relation.  

\begin{acknowledgements}
The observations leading to these results have received funding  from the European Commission's Seventh Framework Programme (FP7/2013-2016)  under grant agreement number 312430 (OPTICON). The authors thank the GAPS observers F.~Borsa, L.~Di Fabrizio, R.~Fares, A.~Fiorenzano, P.~Giacobbe, J.~Maldonado, and G.~Scandariato. The authors thank the CHARA Array, which is funded by the National Science Foundation through NSF grants AST-0606958 and AST-0908253 and by Georgia State University through the College of Arts and Sciences, as well as the W. M. Keck Foundation. This research has made use of the SIMBAD and VIZIER\footnote{Available at http://cdsweb.u- strasbg.fr/} databases at CDS, Strasbourg (France), and of the electronic bibliography maintained by the NASA/ADS system.  WG gratefully acknowledges financial support for this work from the BASAL Centro de Astrofisica y Tecnologias Afines (CATA) PFB-06/2007, and from the Millenium Institute of Astrophysics (MAS) of the Iniciativa Cientifica Milenio del Ministerio de Economia, Fomento y Turismo de Chile, project IC120009. We acknowledge financial support for this work from ECOS-CONICYT grant C13U01. Support from the Polish National Science Center grant MAESTRO 2012/06/A/ST9/00269 is also acknowledged. EP and MR acknowledge financial support from PRIN INAF-2014. NN, PK, AG, and WG acknowledge the support of the French-Chilean exchange program ECOS- Sud/CONICYT (C13U01). The authors acknowledge the support of the French Agence Nationale de la Recherche (ANR), under grant ANR-15-CE31-0012- 01 (project UnlockCepheids) and the financial support from ``Programme National de Physique Stellaire'' (PNPS) of CNRS/INSU, France.
\end{acknowledgements}
\bibliographystyle{aa}  
\bibliography{bibtex_nn} 

\begin{thebibliography}{66}
\expandafter\ifx\csname natexlab\endcsname\relax\def\natexlab#1{#1}\fi

\bibitem[{{Anderson}(2014)}]{anderson14}
{Anderson}, R.~I. 2014, \aap, 566, L10

\bibitem[{{Anderson}(2016)}]{anderson16b}
{Anderson}, R.~I. 2016, \mnras

\bibitem[{{Anderson} {et~al.}(2016){Anderson}, {M{\'e}rand}, {Kervella},
  {Breitfelder}, {LeBouquin}, {Eyer}, {Gallenne}, {Palaversa}, {Semaan},
  {Saesen}, \& {Mowlavi}}]{anderson16}
{Anderson}, R.~I., {M{\'e}rand}, A., {Kervella}, P., {et~al.} 2016, \mnras,
  455, 4231

\bibitem[{{Anderson} {et~al.}(2015){Anderson}, {Sahlmann}, {Holl}, {Eyer},
  {Palaversa}, {Mowlavi}, {S{\"u}veges}, \& {Roelens}}]{anderson15a}
{Anderson}, R.~I., {Sahlmann}, J., {Holl}, B., {et~al.} 2015, \apj, 804, 144

\bibitem[{{Bell} \& {Rodgers}(1964)}]{bell64}
{Bell}, R.~A. \& {Rodgers}, A.~W. 1964, \mnras, 128, 365

\bibitem[{{Benedict} {et~al.}(2002){Benedict}, {McArthur}, {Fredrick},
  {Harrison}, {Slesnick}, {Rhee}, {Patterson}, {Skrutskie}, {Franz},
  {Wasserman}, {Jefferys}, {Nelan}, {van Altena}, {Shelus}, {Hemenway},
  {Duncombe}, {Story}, {Whipple}, \& {Bradley}}]{benedict02}
{Benedict}, G.~F., {McArthur}, B.~E., {Fredrick}, L.~W., {et~al.} 2002, \aj,
  124, 1695

\bibitem[{{Borsa} {et~al.}(2015){Borsa}, {Scandariato}, {Rainer}, {Bignamini},
  {Maggio}, {Poretti}, {Lanza}, {Di Mauro}, {Benatti}, {Biazzo}, {Bonomo},
  {Damasso}, {Esposito}, {Gratton}, {Affer}, {Barbieri}, {Boccato}, {Claudi},
  {Cosentino}, {Covino}, {Desidera}, {Fiorenzano}, {Gandolfi}, {Harutyunyan},
  {Maldonado}, {Micela}, {Molaro}, {Molinari}, {Pagano}, {Pillitteri},
  {Piotto}, {Shkolnik}, {Silvotti}, {Smareglia}, {Southworth}, {Sozzetti}, \&
  {Stelzer}}]{borsa15}
{Borsa}, F., {Scandariato}, G., {Rainer}, M., {et~al.} 2015, \aap, 578, A64

\bibitem[{{Burki} {et~al.}(1982){Burki}, {Mayor}, \& {Benz}}]{burki82}
{Burki}, G., {Mayor}, M., \& {Benz}, W. 1982, \aap, 109, 258

\bibitem[{{Cosentino} {et~al.}(2012){Cosentino}, {Lovis}, {Pepe}, {Collier
  Cameron}, {Latham}, {Molinari}, {Udry}, {Bezawada}, {Black}, {Born},
  {Buchschacher}, {Charbonneau}, {Figueira}, {Fleury}, {Galli}, {Gallie},
  {Gao}, {Ghedina}, {Gonzalez}, {Gonzalez}, {Guerra}, {Henry}, {Horne},
  {Hughes}, {Kelly}, {Lodi}, {Lunney}, {Maire}, {Mayor}, {Micela}, {Ordway},
  {Peacock}, {Phillips}, {Piotto}, {Pollacco}, {Queloz}, {Rice}, {Riverol},
  {Riverol}, {San Juan}, {Sasselov}, {Segransan}, {Sozzetti}, {Sosnowska},
  {Stobie}, {Szentgyorgyi}, {Vick}, \& {Weber}}]{co12}
{Cosentino}, R., {Lovis}, C., {Pepe}, F., {et~al.} 2012, in Society of
  Photo-Optical Instrumentation Engineers (SPIE) Conference Series, Vol. 8446,
  Society of Photo-Optical Instrumentation Engineers (SPIE) Conference Series,
  1

\bibitem[{{Coud{\'e} du Foresto} {et~al.}(1997){Coud{\'e} du Foresto},
  {Ridgway}, \& {Mariotti}}]{foresto97}
{Coud{\'e} du Foresto}, V., {Ridgway}, S., \& {Mariotti}, J.-M. 1997, \aaps,
  121, 379

\bibitem[{{Derekas} {et~al.}(2016){Derekas}, {Plachy}, {Molnar}, {Sodor},
  {Benko}, {Szabados}, {Bognar}, {Csak}, {Szabo}, {Szabo}, \&
  {Pal}}]{derekas16}
{Derekas}, A., {Plachy}, E., {Molnar}, L., {et~al.} 2016, ArXiv e-prints

\bibitem[{{Derekas} {et~al.}(2012){Derekas}, {Szab{\'o}}, {Berdnikov},
  {Szab{\'o}}, {Smolec}, {Kiss}, {Szabados}, {Chadid}, {Evans}, {Kinemuchi},
  {Nemec}, {Seader}, {Smith}, \& {Tenenbaum}}]{derekas12}
{Derekas}, A., {Szab{\'o}}, G.~M., {Berdnikov}, L., {et~al.} 2012, \mnras, 425,
  1312

\bibitem[{{Donati} {et~al.}(1997){Donati}, {Semel}, {Carter}, {Rees}, \&
  {Collier Cameron}}]{donati97}
{Donati}, J.-F., {Semel}, M., {Carter}, B.~D., {Rees}, D.~E., \& {Collier
  Cameron}, A. 1997, \mnras, 291, 658

\bibitem[{{Engle} {et~al.}(2014){Engle}, {Guinan}, {Harper}, {Neilson}, \&
  {Remage Evans}}]{engle14}
{Engle}, S.~G., {Guinan}, E.~F., {Harper}, G.~M., {Neilson}, H.~R., \& {Remage
  Evans}, N. 2014, \apj, 794, 80

\bibitem[{{Evans} {et~al.}(2015){Evans}, {Szab{\'o}}, {Derekas}, {Szabados},
  {Cameron}, {Matthews}, {Sasselov}, {Kuschnig}, {Rowe}, {Guenther}, {Moffat},
  {Rucinski}, \& {Weiss}}]{evans15}
{Evans}, N.~R., {Szab{\'o}}, R., {Derekas}, A., {et~al.} 2015, \mnras, 446,
  4008

\bibitem[{{Fokin}(1991)}]{fokin91}
{Fokin}, A.~B. 1991, \mnras, 250, 258

\bibitem[{{Fouque} \& {Gieren}(1997)}]{fouque97}
{Fouque}, P. \& {Gieren}, W.~P. 1997, \aap, 320, 799

\bibitem[{{Freedman} {et~al.}(2012){Freedman}, {Madore}, {Scowcroft}, {Burns},
  {Monson}, {Persson}, {Seibert}, \& {Rigby}}]{freedman12}
{Freedman}, W.~L., {Madore}, B.~F., {Scowcroft}, V., {et~al.} 2012, \apj, 758,
  24

\bibitem[{{Geroux} \& {Deupree}(2015)}]{geroux15}
{Geroux}, C.~M. \& {Deupree}, R.~G. 2015, \apj, 800, 35

\bibitem[{{Getting}(1934)}]{getting34}
{Getting}, I.~A. 1934, \mnras, 95, 139

\bibitem[{{Gieren} {et~al.}(2005{\natexlab{a}}){Gieren}, {Pietrzynski},
  {Bresolin}, {Kudritzki}, {Minniti}, {Urbaneja}, {Soszynski}, {Storm},
  {Fouque}, {Bono}, {Walker}, \& {Garcia}}]{gieren05_messenger}
{Gieren}, W., {Pietrzynski}, G., {Bresolin}, F., {et~al.} 2005{\natexlab{a}},
  The Messenger, 121, 23

\bibitem[{{Gieren} {et~al.}(2005{\natexlab{b}}){Gieren}, {Storm}, {Barnes},
  {Fouqu{\'e}}, {Pietrzy{\'n}ski}, \& {Kienzle}}]{gieren05}
{Gieren}, W., {Storm}, J., {Barnes}, III, T.~G., {et~al.} 2005{\natexlab{b}},
  \apj, 627, 224

\bibitem[{{Gray} \& {Stevenson}(2007)}]{gray07}
{Gray}, D.~F. \& {Stevenson}, K.~B. 2007, \pasp, 119, 398

\bibitem[{{Groenewegen}(2007)}]{gro07}
{Groenewegen}, M.~A.~T. 2007, \aap, 474, 975

\bibitem[{{Groenewegen}(2013)}]{gro13}
{Groenewegen}, M.~A.~T. 2013, \aap, 550, A70

\bibitem[{{Hadrava} {et~al.}(2009){Hadrava}, {{\v S}lechta}, \& {{\v
  S}koda}}]{hadrava09b}
{Hadrava}, P., {{\v S}lechta}, M., \& {{\v S}koda}, P. 2009, \aap, 507, 397

\bibitem[{{Hertzsprung}(1913)}]{hertzsprung13}
{Hertzsprung}, E. 1913, Astronomische Nachrichten, 196, 201

\bibitem[{{Houdek} \& {Dupret}(2015)}]{houdek15}
{Houdek}, G. \& {Dupret}, M.-A. 2015, Living Reviews in Solar Physics, 12

\bibitem[{{Karp}(1975)}]{karp75}
{Karp}, A.~H. 1975, \apj, 201, 641

\bibitem[{{Kervella} {et~al.}(2004{\natexlab{a}}){Kervella}, {Bersier},
  {Mourard}, {Nardetto}, \& {Coud{\'e} du Foresto}}]{kervella04b}
{Kervella}, P., {Bersier}, D., {Mourard}, D., {Nardetto}, N., \& {Coud{\'e} du
  Foresto}, V. 2004{\natexlab{a}}, \aap, 423, 327

\bibitem[{{Kervella} {et~al.}(2004{\natexlab{b}}){Kervella}, {Fouqu{\'e}},
  {Storm}, {Gieren}, {Bersier}, {Mourard}, {Nardetto}, \& {du Coud{\'e}
  Foresto}}]{kervella04d}
{Kervella}, P., {Fouqu{\'e}}, P., {Storm}, J., {et~al.} 2004{\natexlab{b}},
  \apjl, 604, L113

\bibitem[{{Laney} \& {Joner}(2009)}]{laney09}
{Laney}, C.~D. \& {Joner}, M.~D. 2009, in American Institute of Physics
  Conference Series, Vol. 1170, American Institute of Physics Conference
  Series, ed. J.~A. {Guzik} \& P.~A. {Bradley}, 93--95

\bibitem[{{Leavitt}(1908)}]{leavitt08}
{Leavitt}, H.~S. 1908, Annals of Harvard College Observatory, 60, 87

\bibitem[{{Leavitt} \& {Pickering}(1912)}]{leavitt1912}
{Leavitt}, H.~S. \& {Pickering}, E.~C. 1912, Harvard College Observatory
  Circular, 173, 1

\bibitem[{{Majaess} {et~al.}(2012){Majaess}, {Turner}, \& {Gieren}}]{majaess12}
{Majaess}, D., {Turner}, D., \& {Gieren}, W. 2012, \apj, 747, 145

\bibitem[{{Marengo} {et~al.}(2003){Marengo}, {Karovska}, {Sasselov},
  {Papaliolios}, {Armstrong}, \& {Nordgren}}]{marengo03}
{Marengo}, M., {Karovska}, M., {Sasselov}, D.~D., {et~al.} 2003, \apj, 589, 968

\bibitem[{{Marengo} {et~al.}(2002){Marengo}, {Sasselov}, {Karovska},
  {Papaliolios}, \& {Armstrong}}]{marengo02}
{Marengo}, M., {Sasselov}, D.~D., {Karovska}, M., {Papaliolios}, C., \&
  {Armstrong}, J.~T. 2002, \apj, 567, 1131

\bibitem[{{Merand} {et~al.}(2015){Merand}, {Kervella}, {Breitfelder},
  {Gallenne}, {Coude du Foresto}, {ten Brummelaar}, {McAlister}, {Ridgway},
  {Sturmann}, {Sturmann}, \& {Turner}}]{merand15}
{Merand}, A., {Kervella}, P., {Breitfelder}, J., {et~al.} 2015, ArXiv e-prints

\bibitem[{{M{\'e}rand} {et~al.}(2005){M{\'e}rand}, {Kervella}, {Coud{\'e} du
  Foresto}, {Ridgway}, {Aufdenberg}, {ten Brummelaar}, {Berger}, {Sturmann},
  {Sturmann}, {Turner}, \& {McAlister}}]{merand05}
{M{\'e}rand}, A., {Kervella}, P., {Coud{\'e} du Foresto}, V., {et~al.} 2005,
  \aap, 438, L9

\bibitem[{{Nardetto} {et~al.}(2004){Nardetto}, {Fokin}, {Mourard}, {Mathias},
  {Kervella}, \& {Bersier}}]{nardetto04}
{Nardetto}, N., {Fokin}, A., {Mourard}, D., {et~al.} 2004, \aap, 428, 131

\bibitem[{{Nardetto} {et~al.}(2009){Nardetto}, {Gieren}, {Kervella},
  {Fouqu{\'e}}, {Storm}, {Pietrzynski}, {Mourard}, \& {Queloz}}]{nardetto09}
{Nardetto}, N., {Gieren}, W., {Kervella}, P., {et~al.} 2009, \aap, 502, 951

\bibitem[{{Nardetto} {et~al.}(2013){Nardetto}, {Mathias}, {Fokin},
  {Chapellier}, {Pietrzynski}, {Gieren}, {Graczyk}, \& {Mourard}}]{nardetto13}
{Nardetto}, N., {Mathias}, P., {Fokin}, A., {et~al.} 2013, \aap, 553, A112

\bibitem[{{Nardetto} {et~al.}(2016){Nardetto}, {M{\'e}rand}, {Mourard},
  {Storm}, {Gieren}, {Fouqu{\'e}}, {Gallenne}, {Graczyk}, {Kervella},
  {Neilson}, {Pietrzynski}, {Pilecki}, {Breitfelder}, {Berio}, {Challouf},
  {Clausse}, {Ligi}, {Mathias}, {Meilland}, {Perraut}, {Poretti}, {Rainer},
  {Spang}, {Stee}, {Tallon-Bosc}, \& {ten Brummelaar}}]{nardetto16a}
{Nardetto}, N., {M{\'e}rand}, A., {Mourard}, D., {et~al.} 2016, \aap, 593, A45

\bibitem[{{Nardetto} {et~al.}(2006){Nardetto}, {Mourard}, {Kervella},
  {Mathias}, {M{\'e}rand}, \& {Bersier}}]{nardetto06a}
{Nardetto}, N., {Mourard}, D., {Kervella}, P., {et~al.} 2006, \aap, 453, 309

\bibitem[{{Nardetto} {et~al.}(2007){Nardetto}, {Mourard}, {Mathias}, {Fokin},
  \& {Gillet}}]{nardetto07}
{Nardetto}, N., {Mourard}, D., {Mathias}, P., {Fokin}, A., \& {Gillet}, D.
  2007, \aap, 471, 661

\bibitem[{{Nardetto} {et~al.}(2014{\natexlab{a}}){Nardetto}, {Poretti},
  {Rainer}, {Guiglion}, {Scardia}, {Schmid}, \& {Mathias}}]{nardetto14}
{Nardetto}, N., {Poretti}, E., {Rainer}, M., {et~al.} 2014{\natexlab{a}}, \aap,
  561, A151

\bibitem[{{Nardetto} {et~al.}(2008){Nardetto}, {Stoekl}, {Bersier}, \&
  {Barnes}}]{nardetto08a}
{Nardetto}, N., {Stoekl}, A., {Bersier}, D., \& {Barnes}, T.~G. 2008, \aap,
  489, 1255

\bibitem[{{Nardetto} {et~al.}(2014{\natexlab{b}}){Nardetto}, {Storm}, {Gieren},
  {Pietrzy{\'n}ski}, \& {Poretti}}]{nardetto14b}
{Nardetto}, N., {Storm}, J., {Gieren}, W., {Pietrzy{\'n}ski}, G., \& {Poretti},
  E. 2014{\natexlab{b}}, in IAU Symposium, Vol. 301, Precision
  Asteroseismology, ed. J.~A. {Guzik}, W.~J. {Chaplin}, G.~{Handler}, \&
  A.~{Pigulski}, 145--148

\bibitem[{{Neilson} \& {Ignace}(2014)}]{neilson14}
{Neilson}, H.~R. \& {Ignace}, R. 2014, \aap, 563, L4

\bibitem[{{Neilson} {et~al.}(2012){Neilson}, {Nardetto}, {Ngeow}, {Fouqu{\'e}},
  \& {Storm}}]{neilson12}
{Neilson}, H.~R., {Nardetto}, N., {Ngeow}, C.-C., {Fouqu{\'e}}, P., \& {Storm},
  J. 2012, \aap, 541, A134

\bibitem[{{Ngeow} {et~al.}(2012){Ngeow}, {Neilson}, {Nardetto}, \&
  {Marengo}}]{ngeow12}
{Ngeow}, C.-C., {Neilson}, H.~R., {Nardetto}, N., \& {Marengo}, M. 2012, \aap,
  543, A55

\bibitem[{{Pilecki} {et~al.}(2013){Pilecki}, {Graczyk}, {Pietrzy{\'n}ski},
  {Gieren}, {Thompson}, {Freedman}, {Scowcroft}, {Madore}, {Udalski},
  {Soszy{\'n}ski}, {Konorski}, {Smolec}, {Nardetto}, {Bono}, {Prada Moroni},
  {Storm}, \& {Gallenne}}]{pilecki13}
{Pilecki}, B., {Graczyk}, D., {Pietrzy{\'n}ski}, G., {et~al.} 2013, \mnras,
  436, 953

\bibitem[{{Poretti} {et~al.}(2015){Poretti}, {Le Borgne}, {Rainer}, {Baglin},
  {Benk{\H o}}, {Debosscher}, \& {Weiss}}]{poretti15}
{Poretti}, E., {Le Borgne}, J.~F., {Rainer}, M., {et~al.} 2015, \mnras, 454,
  849

\bibitem[{{Pr{\v s}a} {et~al.}(2016){Pr{\v s}a}, {Harmanec}, {Torres},
  {Mamajek}, {Asplund}, {Capitaine}, {Christensen-Dalsgaard}, {Depagne},
  {Haberreiter}, {Hekker}, {Hilton}, {Kopp}, {Kostov}, {Kurtz}, {Laskar},
  {Mason}, {Milone}, {Montgomery}, {Richards}, {Schmutz}, {Schou}, \&
  {Stewart}}]{prsa16}
{Pr{\v s}a}, A., {Harmanec}, P., {Torres}, G., {et~al.} 2016, \aj, 152, 41

\bibitem[{{Riess} {et~al.}(2011){Riess}, {Macri}, {Casertano}, {Lampeitl},
  {Ferguson}, {Filippenko}, {Jha}, {Li}, \& {Chornock}}]{riess11}
{Riess}, A.~G., {Macri}, L., {Casertano}, S., {et~al.} 2011, \apj, 730, 119

\bibitem[{{Riess} {et~al.}(2016){Riess}, {Macri}, {Hoffmann}, {Scolnic},
  {Casertano}, {Filippenko}, {Tucker}, {Reid}, {Jones}, {Silverman},
  {Chornock}, {Challis}, {Yuan}, {Brown}, \& {Foley}}]{riess16}
{Riess}, A.~G., {Macri}, L.~M., {Hoffmann}, S.~L., {et~al.} 2016, ArXiv
  e-prints

\bibitem[{{Sabbey} {et~al.}(1995){Sabbey}, {Sasselov}, {Fieldus}, {Lester},
  {Venn}, \& {Butler}}]{sabbey95}
{Sabbey}, C.~N., {Sasselov}, D.~D., {Fieldus}, M.~S., {et~al.} 1995, \apj, 446,
  250

\bibitem[{{Sanford}(1956)}]{sanford56}
{Sanford}, R.~F. 1956, \apj, 123, 201

\bibitem[{{Sasselov} \& {Lester}(1990)}]{sasselov90}
{Sasselov}, D.~D. \& {Lester}, J.~B. 1990, \apj, 362, 333

\bibitem[{{Storm} {et~al.}(2011{\natexlab{a}}){Storm}, {Gieren}, {Fouqu{\'e}},
  {Barnes}, {Pietrzy{\'n}ski}, {Nardetto}, {Weber}, {Granzer}, \&
  {Strassmeier}}]{storm11a}
{Storm}, J., {Gieren}, W., {Fouqu{\'e}}, P., {et~al.} 2011{\natexlab{a}}, \aap,
  534, A94

\bibitem[{{Storm} {et~al.}(2011{\natexlab{b}}){Storm}, {Gieren}, {Fouqu{\'e}},
  {Barnes}, {Soszy{\'n}ski}, {Pietrzy{\'n}ski}, {Nardetto}, \&
  {Queloz}}]{storm11b}
{Storm}, J., {Gieren}, W., {Fouqu{\'e}}, P., {et~al.} 2011{\natexlab{b}}, \aap,
  534, A95

\bibitem[{{ten Brummelaar} {et~al.}(2005){ten Brummelaar}, {McAlister},
  {Ridgway}, {Bagnuolo}, {Turner}, {Sturmann}, {Sturmann}, {Berger}, {Ogden},
  {Cadman}, {Hartkopf}, {Hopper}, \& {Shure}}]{ten05}
{ten Brummelaar}, T.~A., {McAlister}, H.~A., {Ridgway}, S.~T., {et~al.} 2005,
  \apj, 628, 453

\bibitem[{{van Hoof} \& {Deurinck}(1952)}]{vanhoof52}
{van Hoof}, A. \& {Deurinck}, R. 1952, \apj, 115, 166

\bibitem[{{van Leeuwen}(2007)}]{leeuw2007b}
{van Leeuwen}, F. 2007, in Astrophysics and Space Science Library, Vol. 350,
  Astrophysics and Space Science Library, ed. {F.~van Leeuwen}

\bibitem[{{van Leeuwen} {et~al.}(2007){van Leeuwen}, {Feast}, {Whitelock}, \&
  {Laney}}]{vl07}
{van Leeuwen}, F., {Feast}, M.~W., {Whitelock}, P.~A., \& {Laney}, C.~D. 2007,
  \mnras, 379, 723

\bibitem[{{Wallerstein} {et~al.}(2015){Wallerstein}, {Albright}, \&
  {Ritchey}}]{wallerstein15}
{Wallerstein}, G., {Albright}, M.~B., \& {Ritchey}, A.~M. 2015, \pasp, 127, 503

\end{thebibliography}
 
\end{document}